\documentclass[aps,prc,preprint,showpacs,floatfix,nofootinbib]{revtex4}

\usepackage{amsmath}
\usepackage{amssymb}
\newcommand{\be}{\begin{equation}}
\newcommand{\ee}{\end{equation}}
\newcommand{\bel}[1]{\be\label{#1}}
\newcommand{\re}[1]{Eq.~(\ref{#1})}
\newcommand{\ds}{\displaystyle}
\newcommand{\hsp}{\hspace*{1pt}}
\newcommand{\hspm}{\hspace*{.5pt}}
\newcommand{\ov}[1]{\overline{#1}}
\usepackage{color}

\usepackage{graphicx}
\usepackage{bm}
\usepackage{hyperref}

\begin{document}
\title{Electromagnetic probes of a pure-glue initial state in nucleus-nucleus collisions at energies available at the CERN Large Hadron Collider}

\author{V.~Vovchenko$^{1,2,3}$, Iu.~A.~Karpenko$^{4,5}$, M.~I.~Gorenstein$^{1,4}$,\\
L.~M.~Satarov$^{1,6}$, I.~N.~Mishustin$^{1,6}$, B.~K\"ampfer$^{7,8}$, and
H.~Stoecker$^{1,2,9}$}
\affiliation{
$^1$\mbox{Frankfurt Institute for Advanced Studies, D-60438 Frankfurt, Germany}\\
$^2$\mbox{Johann Wolfgang Goethe Universit\"at, D-60438 Frankfurt, Germany}\\
$^3$\mbox{Taras Shevchenko National University of Kiev, 03022 Kiev, Ukraine}\\
$^4$\mbox{Bogolyubov Institute for Theoretical Physics, 03680 Kiev, Ukraine}\\
$^5$\mbox{INFN - Sezione di Firenze, I-50019 Sesto Fiorentino (Firenze), Italy}\\
$^6$\mbox{National Research Center ''Kurchatov Institute'', 123182 Moscow, Russia}\\
$^7$\mbox{Helmholtz-Zentrum Dresden-Rossendorf, D-01314 Dresden, Germany}\\
$^8$\mbox{Technische Universit\"at Dresden, Institut f\"ur Theoretische Physik,
D-01062 Dresden, Germany}
$^9$\mbox{GSI Helmholtzzentrum f\"ur Schwerionenforschung GmbH, D-64291 Darmstadt, Germany}
}

\begin{abstract}
Partonic matter produced in the early stage of ultrarelativistic
nucleus-nucleus collisions
is assumed to be composed mainly of gluons, and quarks and antiquarks are produced at later times. To study the implications of such a scenario,
the dynamical evolution of a chemically nonequilibrated
system is described by the ideal (2+1)--dimensional
hydrodynamics with a time dependent (anti)quark fugacity.
The equation of state interpolates linearly between the lattice data
for the pure gluonic matter and the lattice data for the chemically equilibrated quark-gluon plasma.
The spectra and elliptic flows of thermal
dileptons and photons are calculated for central Pb+Pb collisions at the CERN Large Hadron Collider energy of 
$\sqrt{s_{_{\rm NN}}} = 2.76$~TeV.
We test the sensitivity of the results to the choice of equilibration time, including also the case
where the complete chemical equilibrium of partons is reached already at the initial stage.
It is shown that a suppression of quarks at early times leads to a
significant reduction of the yield of the thermal dileptons, but only to a rather modest suppression
of the $p_T$-distribution of direct photons. It is demonstrated that an
enhancement of photon and dilepton elliptic flows might serve as a promising signature of the pure-glue initial state.

\end{abstract}

\maketitle

\section{Introduction}

Strongly interacting matter with extremely high energy density can be created in the laboratory
at the early stages of relativistic nucleus-nucleus (A+A) collisions.  An important physical question
is how the  nonequilibrium initial system of two nucleon counter propagating flows of colliding nuclei
transforms to a state of quarks and gluons in local thermodynamic equilibrium, i.e. to the quark-gluon plasma (QGP). The initial stage
of A+A collisions is presently described by different theoretical models ranging from simple
parton cascades~\cite{Gyulassy,Xu05}, to more sophisticated string-parton models (UrQMD, PHSD, cf.~\cite{UrQMD1,UrQMD2,PHSD}),
color glass condensate~\cite{McLerran}, coherent chromofields~\mbox{\cite{magas-csernai,mishustin-kapusta}}, IP-Glasma~\cite{IPGlasma} etc.
It is usually assumed that strong nonequilibrium effects take place
{only during a} very short proper time interval $\tau_s \sim 1/Q_s$,
where $Q_s\simeq 1\div 2~\textrm{GeV}$ is the so-called gluon saturation
scale~\cite{Gribov}. The idea that the gluonic components of
colliding nucleons dominate in high energy collisions was originally put forward
in Ref.~\cite{Pok-Hove}. It was motivated
by the fact that the perturbative gluon-gluon cross sections are larger than the quark-antiquark
ones.  A two-step equilibration of QGP was proposed in~\cite{raha,shuryak,sinha} assuming that
the gluon thermalization is accomplished already at the early proper time $\sim$ $\tau_s$, while the quark-antiquark chemical
equilibration proceeds until later times $\tau_{\rm th}>\tau_s$. Reference~\cite{Xu05} advocates
that $\tau_{\rm th}=5\div10~\textrm{fm}/c$.
Such a scenario for high energy A+A collisions was considered by several authors, see,
e.g., Refs.~\cite{shuryak,Bir93,BK1,BK2,Tra96,Ell00,Dut02,Gel04,Liu14,Mon14,Moreau2015}.
The {\it pure glue} initial scenario of
Pb+Pb collisions at CERN Large Hadron Collider (LHC)
energies was recently discussed in~Refs.~\cite{Sto16,Sto15}.
The particular aspect of entropy generation in the chemically nonequilibrated QGP has been addressed in~\cite{Vov16}.

In order to highlight possible signatures of the pure-glue initial scenario,
below we describe the evolution of the QGP created in central A+A collisions
using the (2+1)--dimensional boost-invariant hydrodynamics. In our approach
the quark-antiquark fugacity is introduced to describe the QGP %hydrodynamical
evolution
in the absence of the chemical equilibrium.  The main emphasis is
put on electromagnetic probes (thermal photons and dileptons), which may carry an important information
about the deconfined phase. This problem has been repeatedly addressed in the literature, see,
e.g., \cite{BK1,BK2,Gel04,Liu14,Mon14}, however, a definitive conclusion about the role of chemically nonequilibrium evolution
is still missing. The new aspects of the present study include constructing
the equation of state for chemically nonequilibrated QCD matter
via an interpolation of the lattice data, as well as analyzing
the impact of chemical nonequilibrium effects on the dilepton elliptic flow, and demonstrating the importance of the late 'hadronic' stage for the photon spectra.

The paper is organized as follows.
In Sec.~II we formulate the hydrodynamical model used in our calculations.
The equation of state of a chemically nonequilibrated system is constructed by
interpolating the lattice results between the pure gluon and the (2+1)-flavour QCD matter.
In Sec.~III we give some results concerning the space-time evolution
of strongly interacting matter produced in central A+A collisions at LHC energies.
Spectra and elliptic flows of direct photons and thermal dileptons are analysed,
respectively, in Sec.~IV and~V. Our conclusions are given in Sec.~VI.
Appendices A and B provide formulas for photon and dilepton rates, respectively.

\section{\protect Formulation of the model}

\subsection{Equations of motion}
We use a longitudinally boost-invariant (2+1)--dimensional ideal hydrodynamics to
describe the evolution of the net baryon-free matter produced in the high-energy A+A collisions.
The equations of the relativistic hydrodynamics can be written as (\mbox{$\hbar=c=1$})
\bel{hydro}
\frac{\partial\hsp T^{\mu\nu}}{\partial\hsp x^{\nu}}~=~0~,
\ee
where
\bel{tmunu}
T^{\mu\nu}~=~(\varepsilon + P)\hsp u^{\mu}u^{\nu}~-~Pg^{\mu\nu}~
\ee
is the energy-momentum tensor, $u^\mu$ is the four-velocity, $\varepsilon$ and $P$ are the local rest-frame energy density and pressure, respectively, and $g^{\mu\nu}$ is the metric tensor with $g^{00}=1$ in Cartesian coordinates $(t,x,y,z)$, 
with $z$ oriented along the beam axis.
Below we use the curvilinear light-cone 
coordinates $(\tau,x,y,\eta)$, where $\tau = \sqrt{t^2 - z^2}$ is the proper time and \mbox{$\eta=\frac{1}{2} \ln \frac{t+z}{t-z}$}~ is the space-time rapidity.
In the case of the longitudinal boost--invariant (2+1)--dimensional flow one can represent the %components of 
the fluid's four-velocity as~\cite{Kaj86,BK1}
\bel{boost}
u^{\mu}=\gamma_{\perp} (\cosh{\eta}, {\bm v}_{\perp}, \sinh{\eta}),
\ee
where \mbox{${\bm v}_{\perp}$} is the transverse velocity in the symmetry
plane \mbox{$z=0$} and \mbox{$\gamma_{\perp} = (1 - v_{\perp}^2)^{-1/2}$} stands for the
transverse Lorentz factor.
To solve Eq.~(\ref{hydro}) one needs the equation of state (EoS), i.e.,
a relation connecting $P$ and~$\varepsilon$. For chemically nonequilibrated matter considered in this paper,  $P=P\hsp (\varepsilon,\lambda)$, where $\lambda$ is the (anti)quark fugacity. In principle, one should also
solve additional rate equations, defining the evolution of $\lambda$,
as done, e.g., in~\cite{Bir93,BK1,Str94,Mon14}. Instead,
a simple analytic parametrization for~$\lambda$ as a~function of the proper time is used in our study (see the next section).

It is useful to introduce the local proper time $\tau_P$ of a fluid cell element.
Its space-time dependence is determined by the equations
\bel{taup}
u^{\mu} \partial_{\mu} \tau_P ~ = ~ 1~,~~~~~~
\tau_P(\tau_0,x,y,\eta) ~=~ \tau_0 ~,
\ee
where the parameter $\tau_0$ corresponds to initial longitudinal proper time of the hydrodynamic expansion.
Equation~(\ref{taup}) must be solved simultaneously with \re{hydro}.
In general, $\tau_P$ is smaller than the 'global' time $\tau$ due to the presence of non-zero transverse flow.
In the limiting case of the one-dimensional longitudinal Bjorken expansion \cite{Bjo83}, one has $v_{\perp}=0$ and, consequently, $\tau_P=\tau$.

\subsection{\protect Equation of state of chemically nonequilibrium QCD matter \label{seos}}
\begin{figure}[ht]
\centering
\includegraphics[width=0.49\textwidth]{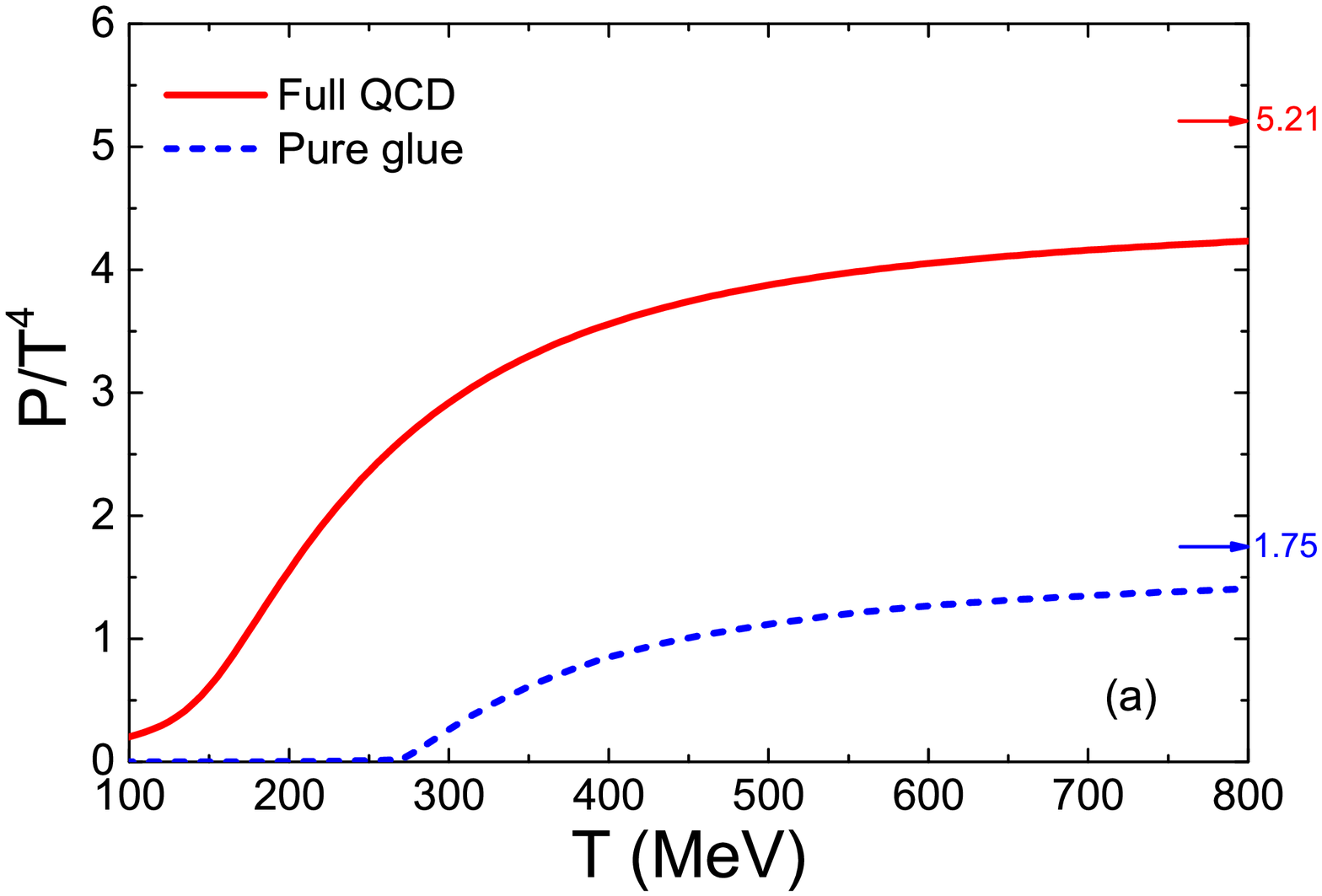}
\includegraphics[width=0.49\textwidth]{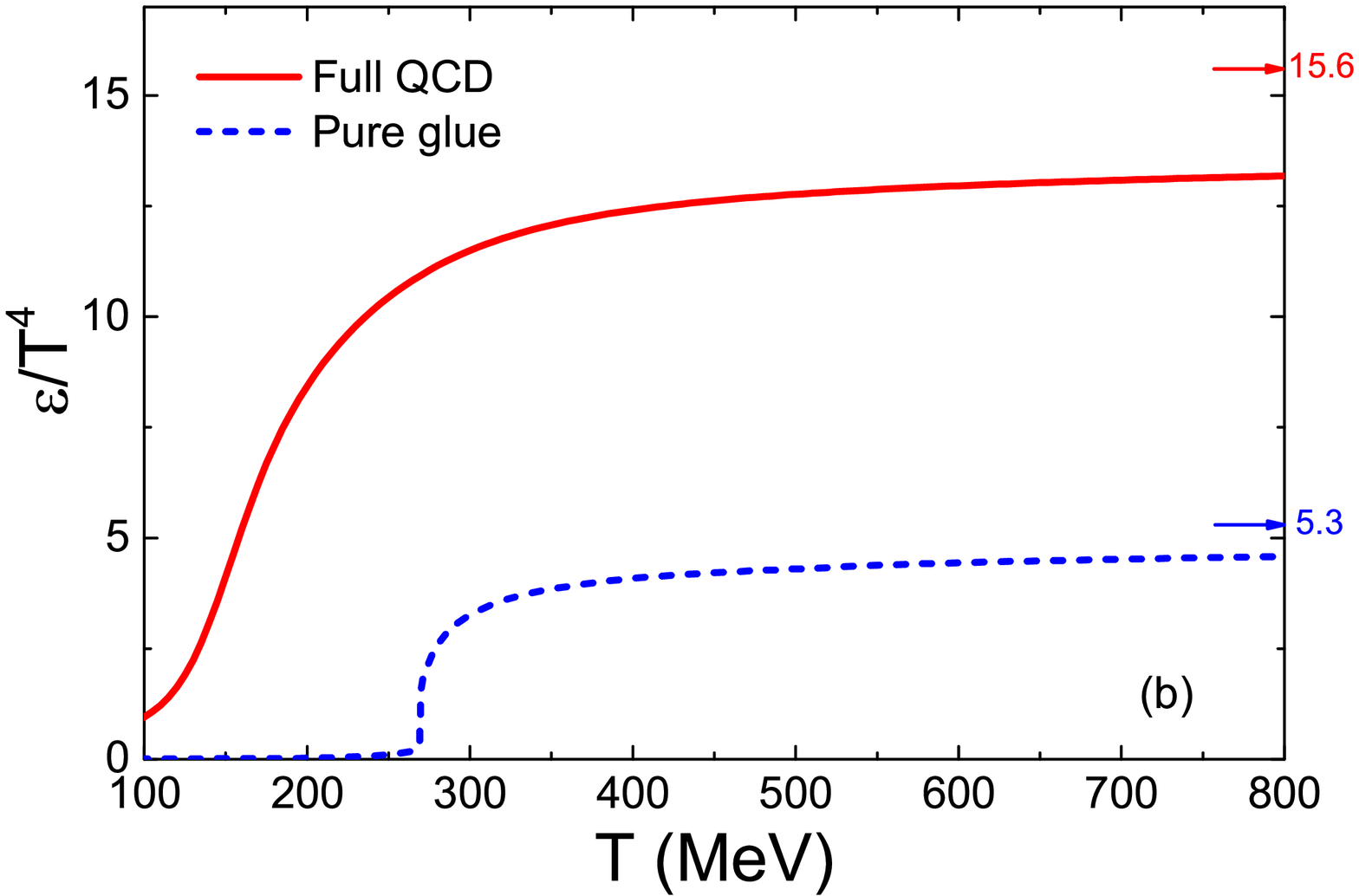}
\caption[]{(Color online)
Temperature dependence of the scaled pressure~(a) and the scaled energy density~(b)
obtained in lattice QCD calculations of Refs.~\cite{WBQCD,WBSU3}.
The solid and dashed lines correspond to the~FQ ($N_f=2+1$) and PG ($N_f=0$)
cases, respectively.
The horizontal arrows indicate
the asymptotic (Stefan-Boltzmann) values of $P/T^4$ and $\varepsilon/T^4$ at large
temperatures.
}\label{fig:latticeEoS}
\end{figure}

We use the lattice {QCD} calculations for the EoS of the {strongly interacting}
matter in two limiting cases: (1)~{the} chemically equilibrated {QCD matter}~\mbox{\cite{WBQCD,HQCD}} and
(2) the SU(3) gluodynamics without (anti)quarks~\cite{Boy95,WBSU3}. {In the following we denote}
these cases as FQ (Full QCD) and PG (Pure Glue), respectively.
The FQ case corresponds to the (2+1)-flavour QCD calculations
which predict the crossover-type
transition at $T\sim 155~\textrm{MeV}$.
The PG calculation provides a first-order phase transition at $T=T_c\simeq 270~\textrm{MeV}$.
The temperature dependencies of the pressure and energy density for FQ and PG scenarios
are exhibited in Fig.~\ref{fig:latticeEoS}.
Larger values of $P$ and~$\varepsilon$ in the FQ calculation appear due to the contribution
of quark-antiquark degrees of freedom. Note the discontinuity of $\varepsilon(T)$
at $T=T_c$ in the PG case. Very small values of $P$ and $\varepsilon$ at $T<T_c$ in the PG matter originate from large masses of glueballs ($M_g\gg T_c$) %representing 
which are
the constituents of the confined phase~\cite{WBSU3}.

The suppression of the quark and antiquark densities
as compared to their equilibrium values at given temperature
is characterized by the (anti)quark fugacity $\lambda$ (for details, see Ref.~\cite{Vov16}).
Generalizing the lattice EoS for the chemically nonequilibrium case
with $\lambda<1$ is not a~straightforward task. We obtain the $P$ and $\varepsilon$ values at fixed $T$ and $\lambda$ by a linear interpolation (LI) between the PG and FQ cases\hsp\footnote
{
For brevity, we denote this equation of state as EoS-LI.
}:
\begin{eqnarray}
P\,(T,\lambda) &=& \lambda \, P_{\hsp\rm FQ}\hsp (T) + (1 - \lambda) \, P_{\hsp\rm PG}\hsp (T)\hsp ,
 \label{eq:pint} \\
\varepsilon\,(T,\lambda) &=& \lambda \, \varepsilon_{\hsp\rm FQ}\hsp (T) + (1 - \lambda) \,
\varepsilon_{\hsp\rm PG}\hsp (T)\hsp .
\label{eq:eint}
\end{eqnarray}

After excluding the temperature variable in Eqs.~(\ref{eq:pint}) and (\ref{eq:eint}), one gets the
relation $P=P(\varepsilon,\lambda)$ which is used in hydrodynamic simulations.
The limits $\lambda=0$ and $\lambda=1$ correspond to the thermodynamic functions of the PG
and FQ matter, respectively. Note that the linear \mbox{$\lambda$--\hsp dependence} of $P$ and $\varepsilon$ is a characteristic
feature of the ideal gas of massless gluons and (anti)quarks studied in Refs.~\cite{Sto15,Vov16}.

\begin{figure}[ht]
\centering
\includegraphics[width=0.65\textwidth]{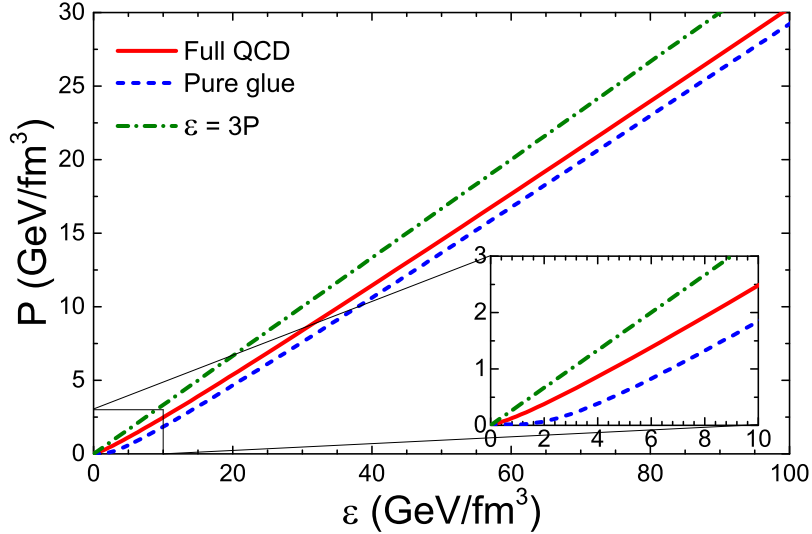}
\caption[]{(Color online)
Pressure as a function of energy density in FQ (solid curve) and PG (dashed curve) cases obtained in lattice calculations~\cite{WBQCD,WBSU3}. Additionally, the $P=\varepsilon/3$ dependence for the ultrarelativistic ideal gas is shown by the dash-dotted line.
The inset zooms into the region of smaller energy densities.
}\label{fig:lattice-p-e}
\end{figure}

It is interesting that the $\varepsilon$--\hsp dependence of the pressure needed for the hydrodynamical calculations appears to be rather similar in PG and FQ matter. This is shown in
Fig.~\ref{fig:lattice-p-e}. The pressure values corresponding to
EoS-LI will change from $P=P_{\,\rm PG}(\varepsilon)$ at the initial stage of the
A+A collision to {$P=P_{\,\rm FQ}(\varepsilon)$} during the later
stage of chemical equilibration\hsp\footnote
{
Possible supercooling phenomena may change this behavior.
}.
As follows from Fig.~\ref{fig:lattice-p-e}, both equations of state show an almost linear $P(\varepsilon)$ dependence in the considered energy density range, but they are both softer than the EoS of the ideal gas of massless partons.

Using Eqs.~(\ref{eq:pint}) and (\ref{eq:eint}) and basic thermodynamic identities,
one can calculate the total density of (anti)quarks $n_q$ and the entropy density $s$\hsp .
The following relations are obtained
\begin{eqnarray}
n_{\hsp q}\,(T,\lambda) &=& \frac{\lambda}{T} \, ( P_{\hsp\rm FQ}-P_{\hsp\rm PG} ),
\label{eq:nqint} \\
s\,(T,\lambda) &=& \lambda \, s_{\hsp\rm FQ} (T) + (1 - \lambda) \, s_{\hsp\rm PG} (T)
- n_{\hspm q} (T,\lambda)\hsp\ln{\lambda}\hsp .
\label{eq:sint}
\end{eqnarray}

\begin{figure}[ht]
\centering
\includegraphics[width=0.49\textwidth]{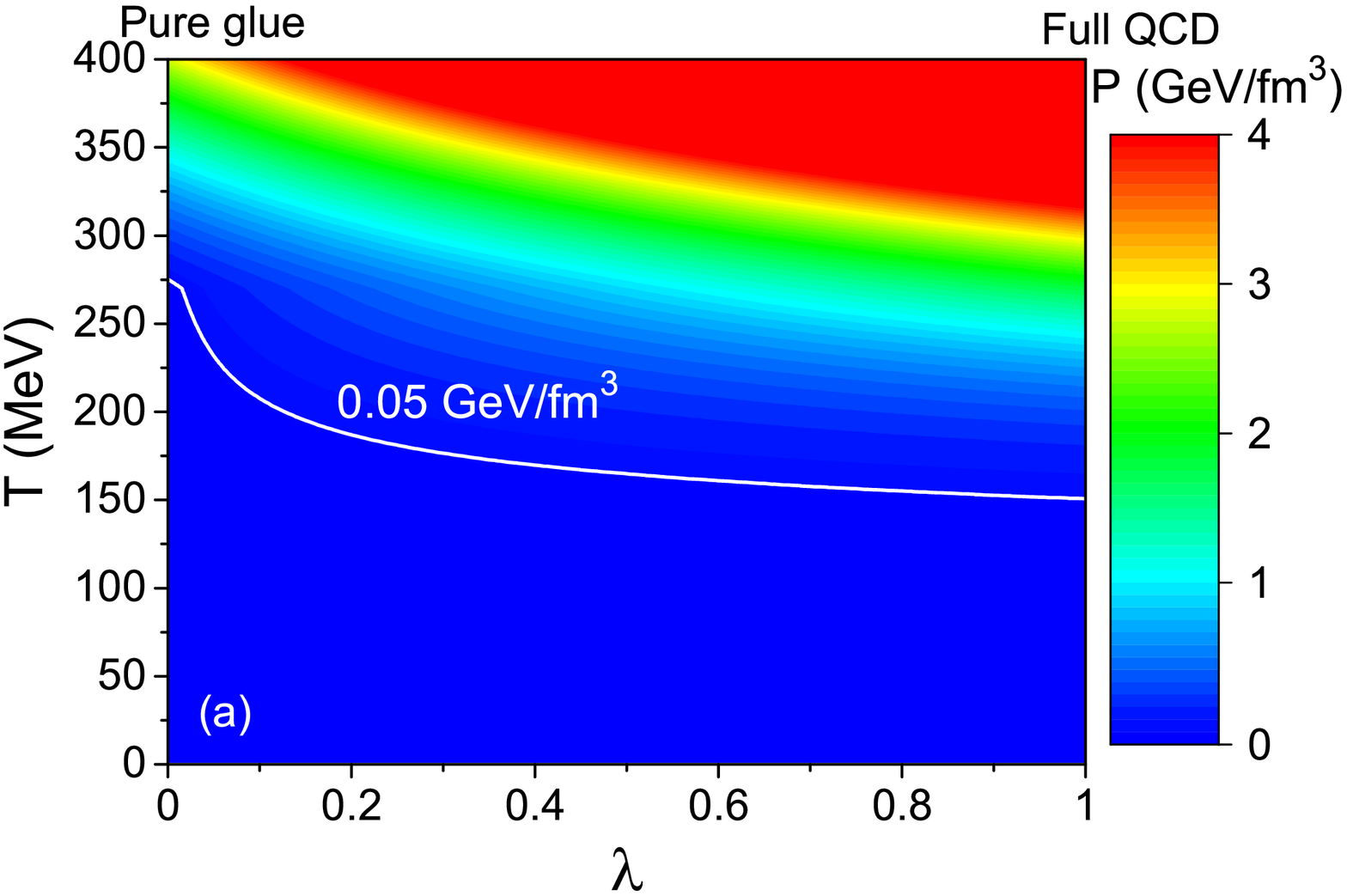}
\includegraphics[width=0.49\textwidth]{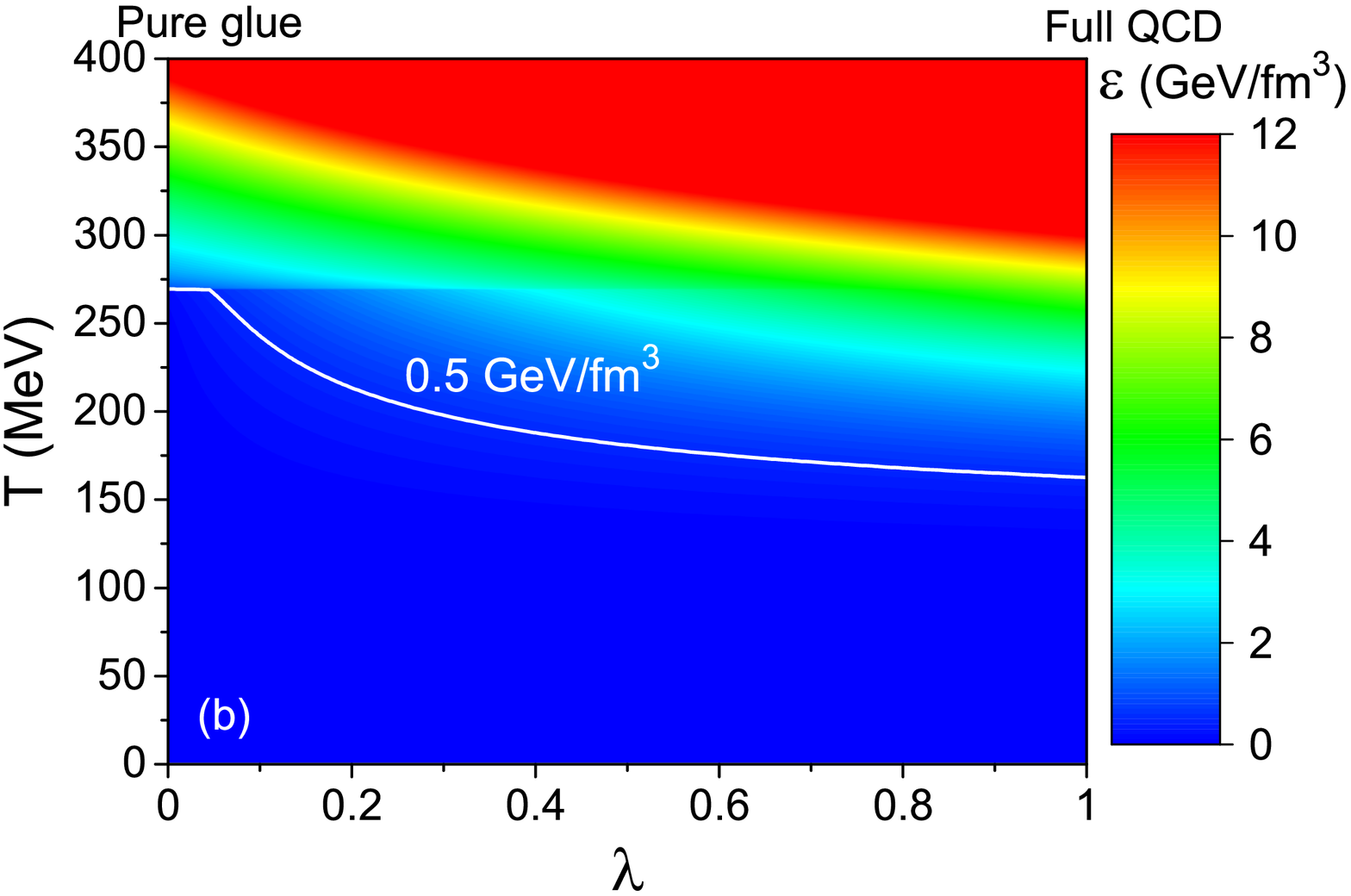}
\caption[]{(Color online)
Contour plots of pressure (a) and energy density (b) {for} chemically non-equilibrated QCD calculated
from Eqs.~\eqref{eq:pint} and \eqref{eq:eint}. The white lines show contours
\mbox{$P=0.05~\textrm{GeV/fm}^3$}~(a) and $\varepsilon=0.5~\textrm{GeV/fm}^3$ (b).
}
\label{fig:PD}
\end{figure}

The two-dimensional plots of $P$ and $\varepsilon$
for the chemically nonequilibrated QCD are shown in Fig.~\ref{fig:PD}. The EoS-LI
contains the first-order phase transition at $T_c=270$~MeV. The latent heat of this transition depends
on $\lambda$, and it goes to zero at~$\lambda\to 1$\hsp .

Below we assume that at $\tau=\tau_0$ the initial (anti)quark
densities vanish in all cells and gluons are in thermal and chemical equilibrium.
Similarly to Refs.~\cite{Sto15,Vov16} we postulate that~$\lambda$ is an
explicit function of the local proper time $\tau_P$ which increases from $\lambda=0$ at $\tau_P=\tau_0$ to $\lambda=1$ at $\tau_P-\tau_0\rightarrow \infty$. The following simple parametrization is used:
\bel{lambda}
\lambda(\tau_P) = 1 - \exp\left( \frac{\tau_0-\tau_P}{\tau_*} \right),
\ee
where $\tau_*$ is a model parameter characterizing the quark chemical equilibration time.
{There are different} estimates for $\tau_*$ {in the literature} ranging from
\mbox{$\tau_*\sim 1~\textrm{fm}/c$~\cite{Rug15}} to \mbox{$\tau_*\sim 5~\textrm{fm}/c$~\cite{Xu05}}.
Note that $\tau_*=0$ corresponds to the instantaneous chemical equilibration of quarks and gluons.

In our calculations we assume that gluons are always in thermal and chemical equilibrium immediately from the beginning of the hydro expansion. This assumption can be relaxed by modeling the chemical non-equilibrium of gluons by using the time dependent gluon fugacity $\lambda_g$, 
with a different (smaller) relaxation time $\tau_g$ for gluons compared to quarks. 
The calculations can be made even more
realistic by introducing additional rate equations describing the
space-time evolution of quark and gluon densities. 
However, this would require some new assumptions. 
In particular, the introduction of the chemical non-equilibrium for gluons would  require modifications to the equation of state.
Calculations employing the rate equations~\cite{Ell00,Gel04,Rug15}, as well as those employing the microscopic parton cascade~\cite{Sto16}, indicate that the time evolution of the gluon fugacity is not completely trivial, and may even be non-monotonic.
For our analysis it is most important that, at the early stages, the gluon fugacity is still significantly larger than the (anti)quark fugacity. 
Thus, in the present work we only consider the undersaturation of quarks, but not of gluons.

\subsection{Initial conditions \label{sinc}}
We consider Pb+Pb collisions at the LHC
with center-of-mass ({c.\hsp m.}) energy per nucleon pair $\sqrt{s_{NN}} = 2.76$~TeV.
In our calculations we {choose} $\tau_0 = 0.1$~fm/$c$ as the initial time of {the hydrodynamic} evolution.
It is assumed that there is no initial
transverse flow, i.e., \mbox{$\bm{v}_{\perp}(\tau_0,x,y)=0$}, and the initial energy density profile
is proportional to the linear combination of the transverse distributions of wounded nucleons and of binary collisions taken from the
event-averaged Monte Carlo Glauber model as implemented in the GLISSANDO code~\cite{GLISSANDO}.
The coefficient of proportionality in the initial $\varepsilon$-profile
is fixed to reproduce the {observed} hadron spectra {within} the simulation assuming {\textit{chemical
equilibrium} with the full QCD EoS for a}~given centrality interval~(see Ref.~\cite{KarpenkoLHC} for details). We use the same initial energy density profile in the present calculations for the chemical nonequilibrium case.

It is also assumed that initially the fugacity $\lambda$ of (anti)quarks is zero, i.e. the initial state is purely gluonic. In our model this is realized by setting the initial local proper time of each fluid element equal to $\tau_0$, i.e. 
$\tau_P(\tau_0, x, y, \eta) = \tau_0$. 

\section{Numerical results of hydrodynamic simulations}

Equations (\ref{hydro}) and (\ref{taup}) are solved using the (2+1)--dimensional
version of the vHLLE hydro code~\cite{vHLLE}.
The EoS tables $P=P\hsp (\varepsilon,\lambda)$ for hydrodynamic
simulations were prepared as described in Sec.~\ref{seos}\hsp . We consider the 0\hspm --20\hsp\%
and 20\hspm --40\hsp\% central Pb+Pb collisions.
\begin{figure}[htb!]
\centering
\includegraphics[width=0.65\textwidth]{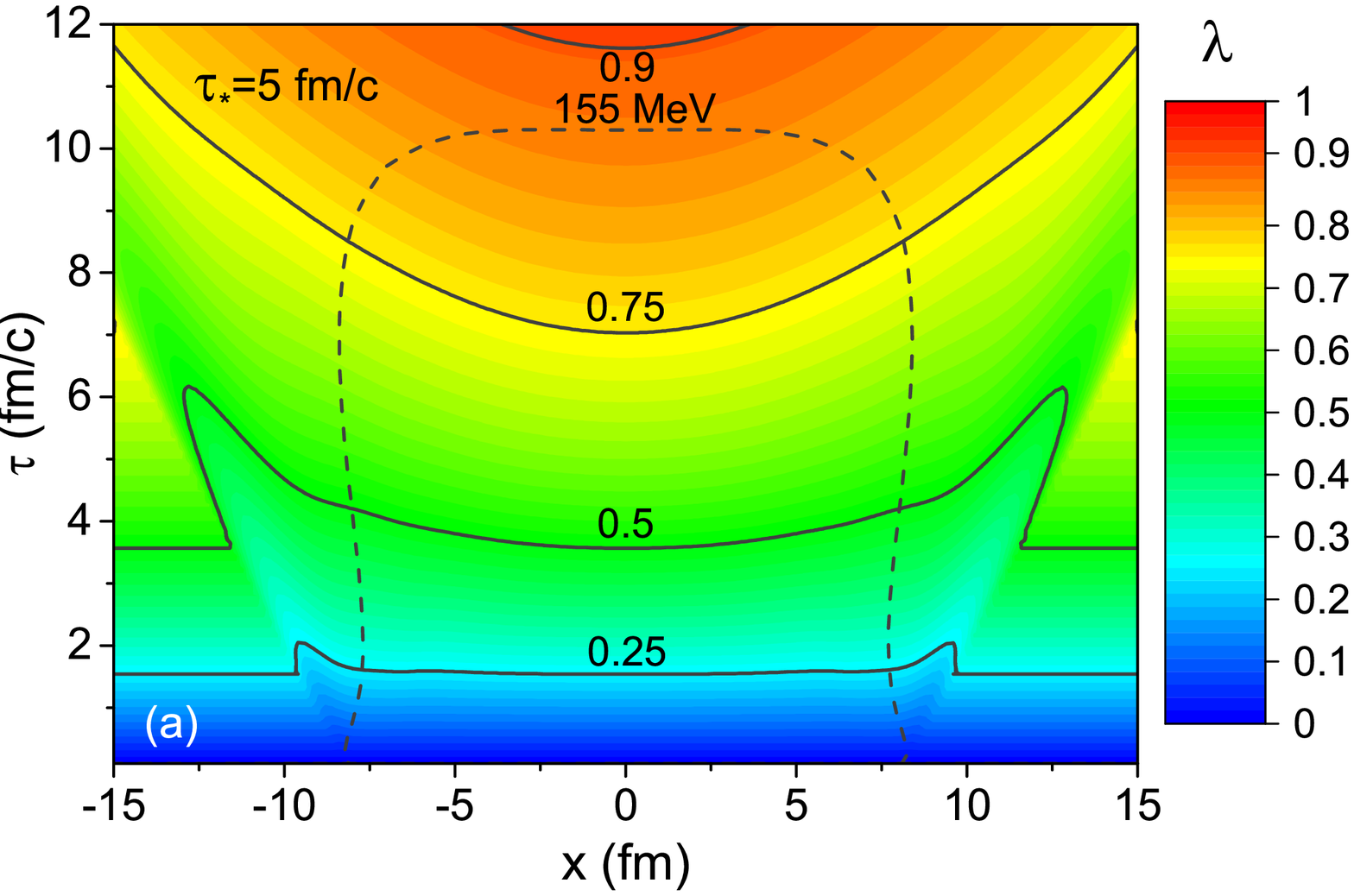}
\includegraphics[width=0.65\textwidth]{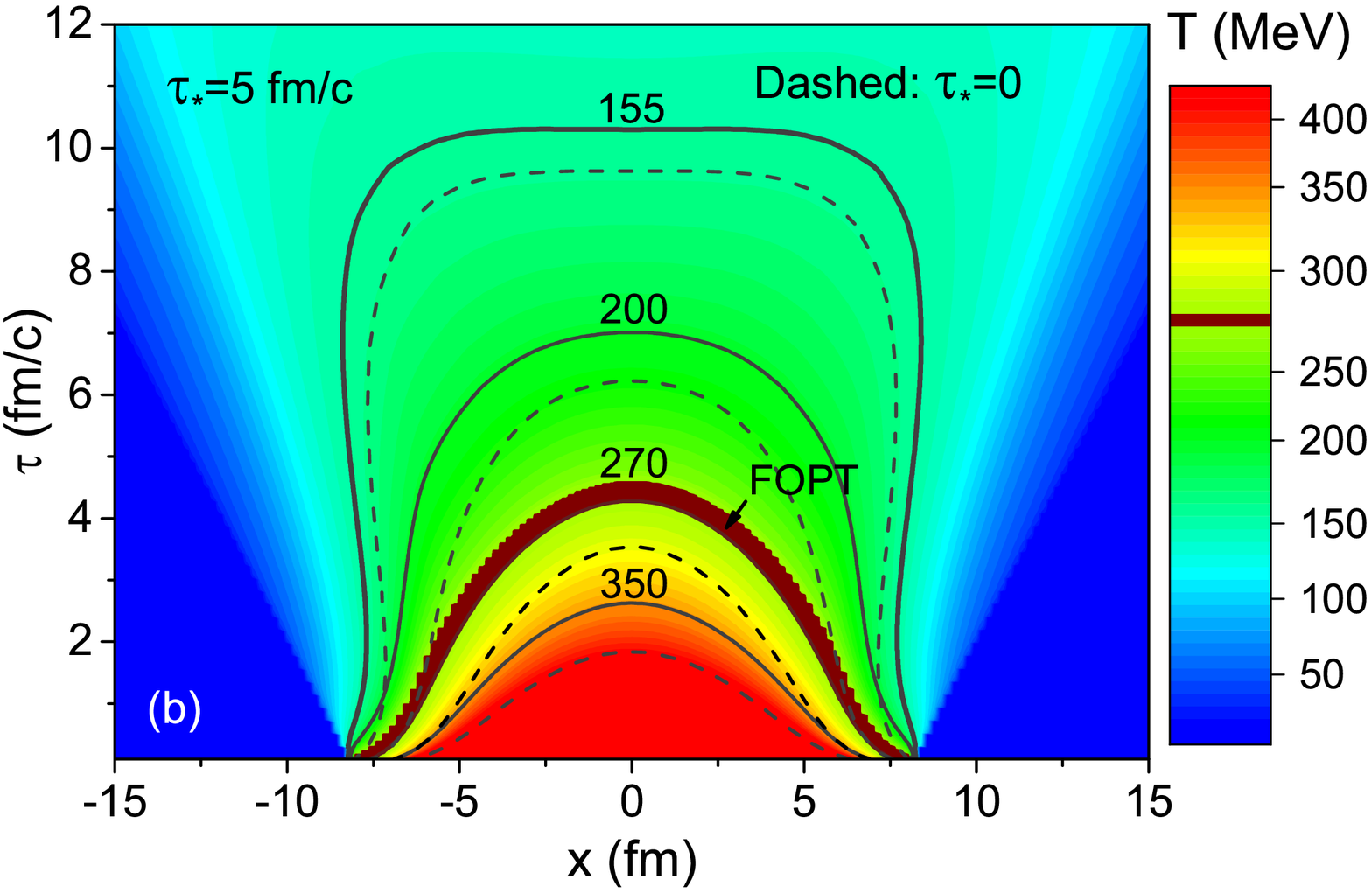}
\caption[]{(Color online)
Contour plots of the {quark fugacity (a) and temperature (b)} in the $x-\tau$ plane for {the 0--20\hspm\% most}
central Pb+Pb collisions at \mbox{$\sqrt{s_{NN}} = 2.76$~TeV}.
The solid curves show {contours of~$\lambda$ and $T$ (in units of MeV)}.
The dashed line {in (a)} corresponds to the isotherm $T=155~\textrm{MeV}$.
The~dark region labeled by FOPT corresponds to the mixed-phase region of the first-order phase
transition at \mbox{$T=T_c\simeq~270~\textrm{MeV}$}. The dashed curves in (b) depict isotherms calculated for
equilibrium matter with $\lambda=1$.
}\label{fig:T}
\end{figure}

The density plot of the quark fugacity $\lambda$ in the $x-\tau$ plane
is given in~Fig.~\ref{fig:T}\hsp a. The dashed line
shows the isotherm $T=155~\textrm{MeV}$ which presumably corresponds to the hadronization
hypersurface.
One can see that typical lifetimes of the deconfined
phase in the considered reaction do not exceed $10~\textrm{fm}/c$\hsp.
In Fig.~\ref{fig:T}\hsp a one observes that deviations from chemical equilibrium {(\mbox{$\lambda\lesssim 0.9$})}
survive up to the hadronization {stage}.
As discussed in Ref.~\cite{Vov16} this may lead to a suppression
of (anti)baryon-to-pion ratios observed~\cite{Abe13} for the considered reaction.
%The possible suppression of hadron observables in pure glue scenario at lower collision energy was also reported in Ref.~\cite{Moreau2015}.
Note that $\lambda$ evolves with $\tau$ in Fig.~\ref{fig:T}\hsp a even in the large $x$ regions where there is virtually no matter, which results from applying the Eq.~\eqref{lambda}. In reality, of course, the values of $\lambda$ for very dilute and cold  fluid elements
%its values there are 
are irrelevant and should be ignored.

Figure ~\ref{fig:T}\hsp b shows the density plot of
the temperature in the coordinates $(x,\tau)$.
The solid and dashed curves correspond to
$\tau_*=5~\textrm{fm}/c$ and $\tau_* = 0$, respectively.
One {can see} that the chemically undersaturated matter is hotter as compared to
the equilibrium case (\mbox{$\lambda=1$})\hsp\footnote
{
Note that in both cases we take the same profile of the energy density at $\tau=\tau_0$.
}. This is a consequence of reduced
number of degrees of freedom in such a medium. According to Fig.~\ref{fig:T}\hsp b,
typical lifetimes of the mixed phase are rather short, they do not
exceed $0.5~\textrm{fm}/c$\hsp . This is at variance with calculations in the (1+1) dimensional hydrodynamics
which predict~\cite{Sat07} much larger lifetimes of the mixed phase within the chemically equilibrated bag model. Therefore, the account of transverse expansion is rather important.

\begin{figure}[ht!]
\centering
\includegraphics[width=0.65\textwidth]{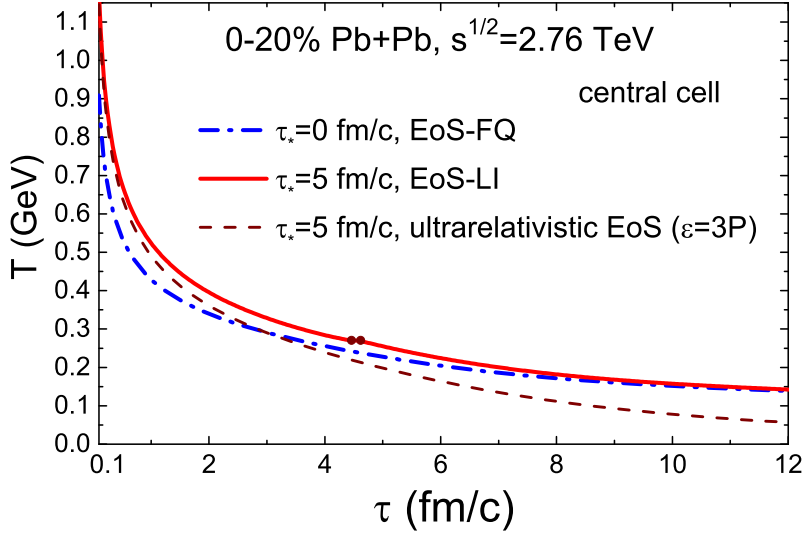}
\caption[]{(Color online)
Temperature in the central cell as a function of proper time
for \mbox{{the 0--20\hspm\%}}~central Pb+Pb collisions at \mbox{$\sqrt{s_{NN}} = 2.76$~TeV}. The dashed and solid curves
are calculated for $\tau_* = 0$ and $\tau_* = 5~\textrm{fm}/c$, respectively. The short section between filled dots on the solid
curve corresponds to mixed-phase states of the confinement phase transition.
The dash-dotted curve is for $\tau_* = 5~\textrm{fm}/c$ assuming the ideal gas EoS.
}\label{fig:Tcentral}
\end{figure}

Figure \ref{fig:Tcentral} shows the evolution of the temperature in the central cell $(x,y,z)=0$ for
$\tau_*=0$ and $\tau_*=5~\textrm{fm}/c$\hsp . In the second case, $\tau_* = 5$~fm/$c$ we compare the calculations for EoS-LI (solid line)
and for the ideal gas of massless partons~\cite{Vov16} (dashed line). One can see
significant differences between these two calculations at late times.

It is evident that the entropy will grow in the course of chemical equilibration.
This was demonstrated in Ref.~\cite{Vov16} within the purely longitudinal Bjorken hydrodynamics. Here we present
a similar analysis within the (2+1)--dimensional hydrodynamical model.
The total amount of entropy flowing through a space-time hypersurface $\sigma^{\hsp\mu}$ can be evaluated as
\cite{Sat07}:
\bel{tente}
S=\int d\hsp\sigma^{\hsp\mu}u_{\mu} s\,.
\ee
Here $s$ is the entropy density and $d\hsp\sigma^{\hsp\mu}$ is the element
of {a} space-time hypersurface which we choose below\hsp\footnote
{
In the case of chemical equilibrium $S$ is constant and does not depend on the choice
of a hypersurface.
}
as the surface of constant proper \mbox{time~$\tau$}. Using~\re{boost}, one can show that
$d\hspm\sigma^{\hspm\mu}u_\mu=\gamma_\perp\tau d^{\hsp 2}x_\perp d\hspm\eta$
for such a hypersurface.
Substituting this relation into~\re{tente} leads to the following expression for the total entropy
per unit space-time \mbox{rapidity} in \mbox{the~(2+1)\hspm --\hsp dimensional} hydrodynamics
\bel{dsdet}
\frac{dS\,(\tau)}{d\hsp\eta}=\tau\int d^{\hsp 2}x_\perp\hsp\gamma_\perp(\tau,\bm{x}_\perp)\hsp s\hsp (\tau,\bm{x}_\perp)\,.
\ee

\vspace*{2mm}
\begin{figure}[htb!]
\centering
\includegraphics[width=0.65\textwidth]{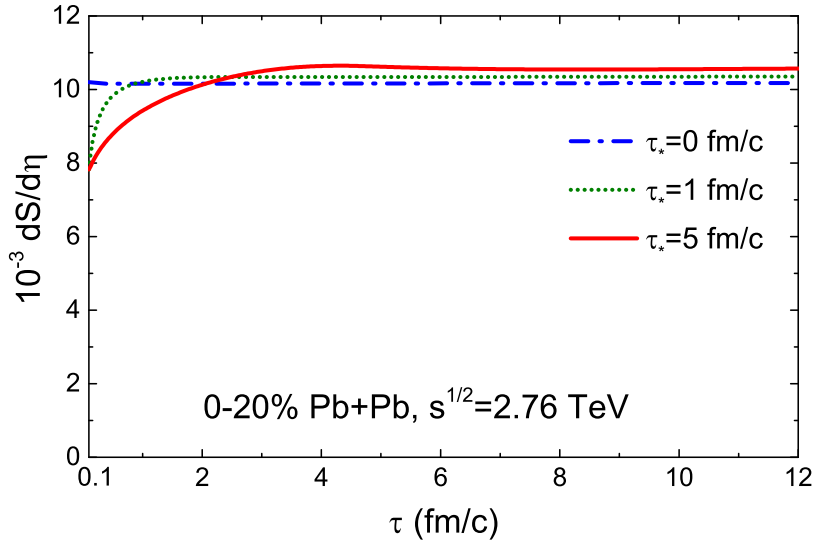}
\caption[]{(Color online)
Total entropy per unit space-time rapidity as a function of proper time~$\tau$ for the
{0--20\hspm\%} central Pb+Pb collisions at $\sqrt{s_{NN}} = 2.76$~TeV. The dash-dotted, dotted and solid curves correspond
to the parameters $\tau_* = 0\hspm{, 1}$ {and} $5~\textrm{fm}/c$\hsp , respectively.
}\label{fig:dSdeta}
\end{figure}
The results of the entropy calculations for the {0--20\hspm\%} central Pb+Pb collisions
are shown in~Fig.~\ref{fig:dSdeta}.
At $\tau_* = 5~\textrm{fm}/c$ the relative increase of the entropy
is about {30\hspm\%}. Approximately the same {relative increase} has been
obtained in \cite{Vov16} within a one-dimensional Bjorken-like
calculation for the ideal gas EoS. Note, that more consistent calculations
for nonzero $\tau_*$ would
require renormalizing the initial energy density profiles to
obtain the same final pion multiplicities as in the equilibrium case. The asymptotic values
of $dS/d\eta$ for {different} choices of $\tau_*$ in Fig.~{\ref{fig:dSdeta}} will be
then the same.

\section{Direct photon emission\label{dpes}}

The emission of direct\hspm\footnote
{
By direct photons we denote the 'non-cocktail' photons i.e.~those which are not produced in decays of
$\pi^0,\eta,\rho,\eta^\prime$, and $\phi$ mesons in the final stage of the reaction.
}
photons from expanding matter created in relativistic A+A collisions
has several components~\cite{Paq15,Lin16}: a)~'prompt' photons from binary collisions of initial nucleons, b)~'thermal' photons
from the high-temperature deconfined phase, c)~direct photons from {the} low-temperature hadronic phase.
The contribution of prompt photons becomes dominant at large transverse momenta.
As we will see below, this greatly reduces the~sensitivity
of photon $p_{\hsp T}$-spectra to chemical nonequilibrium effects.
However, the situ\-ation with transverse flows of photons
is different because of low azimuthal anisotropy of prompt photons.
Note that the \mbox{ALICE} experiments~\cite{Loh13} reveal
large elliptic flows of direct photons, which still can not
be explained in the chemically equilibrium scenario~\cite{Paq15}.

Within the leading order approximation in the strong coupling constant,
the following sources of thermal photon production in the deconfined matter are dominant~\cite{Arn01}:\\
1) QCD Compton scattering (\mbox{$A+g\to A+\gamma$}, where $A=q,\ov{q}$),\\
2) quark-antiquark annihilation (\mbox{$q+\ov{q}\to g+\gamma$}),\\
3) bremsstrahlung reactions (\mbox{$A+B\to A+B+\gamma$}, where $A=q,\ov{q}$ and $B=q,\ov{q},g$),\\
4) 'off-shell' $q\ov{q}$--annihilation with rescatterings of (anti)quark on another parton in the initial state\hsp\footnote
{
According to Ref.~\cite{Ghi13}, the next-to-leading order corrections to the rate of photon production
in equilibrium QGP do not exceed 20\%.
}. It is clear that photons can not be produced in a pure glue matter without
charged (anti)quark partons.

Let us consider the invariant photon production rate (PPR) in the chemically undersaturated quark-gluon plasma (uQGP)
with the temperature $T$ and the quark fugacity $\lambda$. Below we denote this quantity as $\Gamma (\widetilde{E},T,\lambda)$,
where $\widetilde{E}$ is the photon energy in the rest frame of the fluid element. The limiting case of complete
chemical equilibrium ($\lambda=1$) is considered in Appendix~\ref{app-A}. We use the analytic
parametrization {for} $\Gamma (\widetilde{E},T)=\Gamma (\widetilde{E},T,1)$ suggested in~Ref.~\cite{Arn01}.
Equations (\ref{rat12})--(\ref{cexp}) give the explicit expressions for $\Gamma_i(\widetilde{E},T)$, which are
the PPR of processes $i=1,2$ in the chemically equilibrated QGP.

\begin{figure*}[hbt!]
\centering
\includegraphics[width=0.49\textwidth]{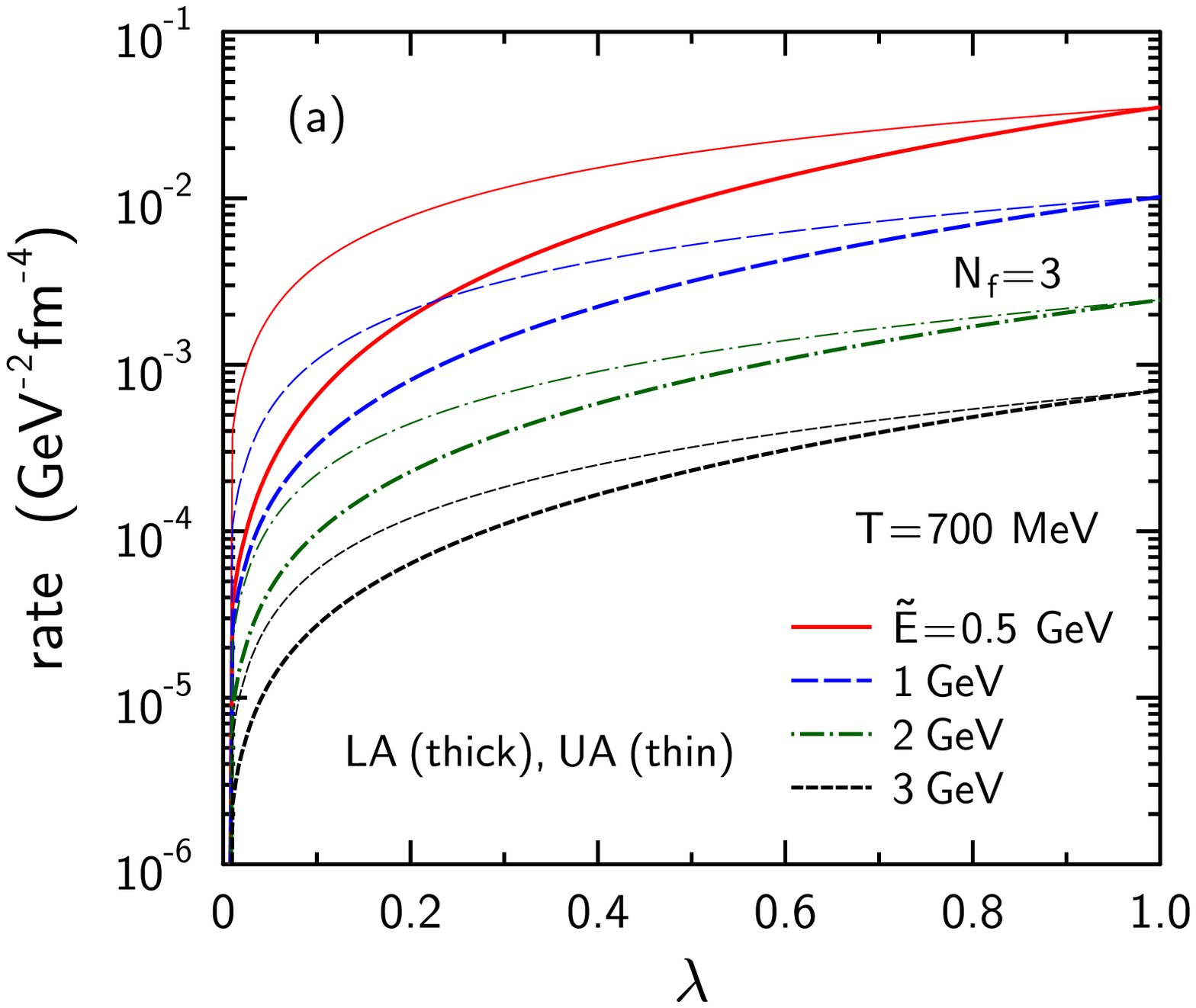}
\includegraphics[width=0.49\textwidth]{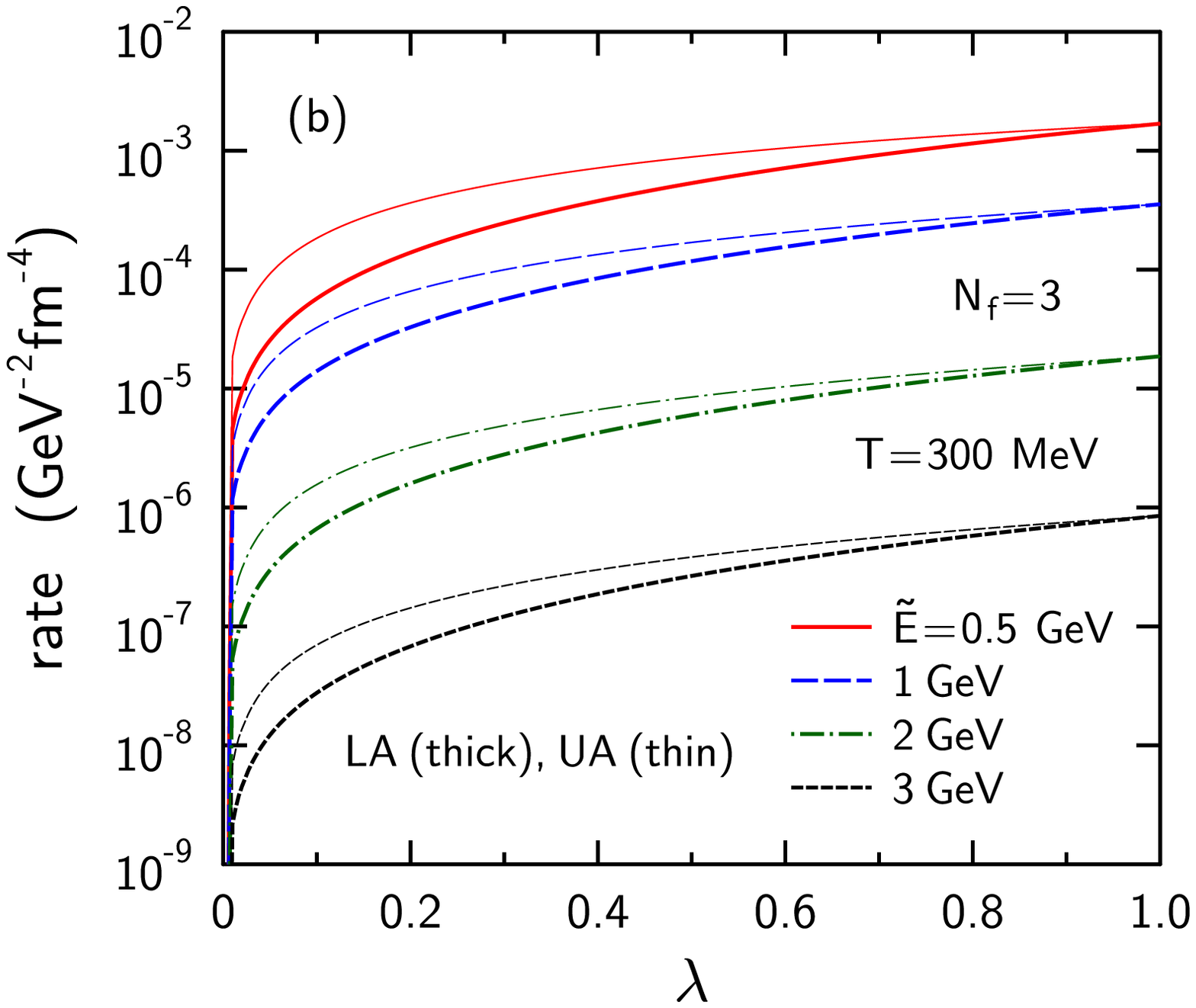}
\caption[]{(Color online)
{Thermal} photon production rates in uQGP as functions of quark fugacity~$\lambda$
at temperature{s} $T=700$~(a)~and $300$~(b)~MeV for different values {of} rest-frame
photon energy~$\widetilde{E}$\hspm . Thick and thin lines are calculated by using~\re{uqgp1} and~(\ref{uqgp2}), respectively.}
\label{fig:lam-ppr}
\end{figure*}

To calculate PPR {for arbitrary $\lambda$} we introduce the additional suppression
factor $\lambda$ for each quark and {antiquark} \mbox{\cite{Sto15,Vov16}} in initial states
of {the processes} 1-4. In particular, {the rates} of the processes 1 and 2 will be suppressed
by the factors~$\lambda$ and $\lambda^2$, respectively.
An analogous procedure for {the} processes~3 and 4 is not {trivial}, as the contribution of partons \mbox{$B=g$}
is not suppressed as compared to \mbox{$B=q,\ov{q}$}\hspm . Similar to Ref.~\cite{Liu14}, we apply two different approximations for
$\Gamma (\widetilde{E},T,\lambda)$:
\begin{eqnarray}
\label{uqgp1}
\textrm{LA:}~~~&&\Gamma\hsp (\widetilde{E},T,\lambda)=\lambda\hsp\Gamma_1(\widetilde{E},T)+
\lambda^2\left[\hsp\Gamma\hsp (\widetilde{E},T)-\Gamma_1(\widetilde{E},T)\hsp\right],\\
\label{uqgp2}
\textrm{UA:}~~&&\Gamma\hsp (\widetilde{E},T,\lambda)=\lambda^2\hsp\Gamma_2(\widetilde{E},T)+
\lambda\left[\hsp\Gamma\hsp (\widetilde{E},T)-\Gamma_2(\widetilde{E},T)\hsp\right],
\end{eqnarray}
where {$\Gamma\hsp (\widetilde{E},T)$} is calculated by using Eqs.~(\ref{ramy1})--(\ref{ramy4}).
It is clear that the approximation LA (UA) underestimates (overestimates) the ``exact''
photon production rate in uQGP. The results of PPR calculations using Eqs.~\mbox{(\ref{uqgp1}) and (\ref{uqgp2})} are shown
in Fig.~\ref{fig:lam-ppr} {for several values of $\widetilde{E}$ and $T$}. One can see that the difference between the parametrizations LA and UA increases with $T$. Note that large temperatures correspond to early stages of a heavy--ion collision,
when the values of $\lambda$ are rather small in the pure glue initial scenario~\cite{Sto15}.

In our case of a boost invariant (2+1)\hspm --\hsp dimensional expansion the
invariant yield of thermal photons is calculated as
\bel{photsp}
\frac{d N_{\gamma}^{\hsp\rm th}}{d^{\hsp 2} p_{\hsp T} dY} =
\int\hspace*{-3pt}d^{\hspace*{1.5pt}2}\hspace*{-1pt}x_T\hspace*{-3pt}\int\limits_{\tau_0}^{+\infty} \hspace*{-3pt}d\hspm\tau\hsp\tau
\hspace*{-3pt}\int\limits_{-\infty}^{+\infty} \hspace*{-3pt}d\hsp\eta \, \Gamma\hsp (\widetilde{E}, T, \lambda)\hsp \theta\hsp (T-T_f)\,,
\ee
where {$p_{\hsp T}$ is the transverse momentum of the photon, $Y$ is its longitudinal rapidity,} $\widetilde{E}=\gamma_{\perp}\,p_{\hsp T}\hsp \big[\hspm\cosh(Y-\eta)-v_x\cos\varphi-v_y\sin\varphi\hsp \big]$
($\varphi$ is the angle between
$\bm{p}_{\,T}$ and the reaction plane), \mbox{$\theta\,(x)=\big[\hspm 1+\textrm{sgn}\hsp (x)\hsp\big]/2$}, and $T_f$ is 
the minimum temperature,
i.e. radiation from fluid cells with $T>T_f$ is considered only.
\begin{figure*}[hbt!]
\centering
\includegraphics[width=0.65\textwidth]{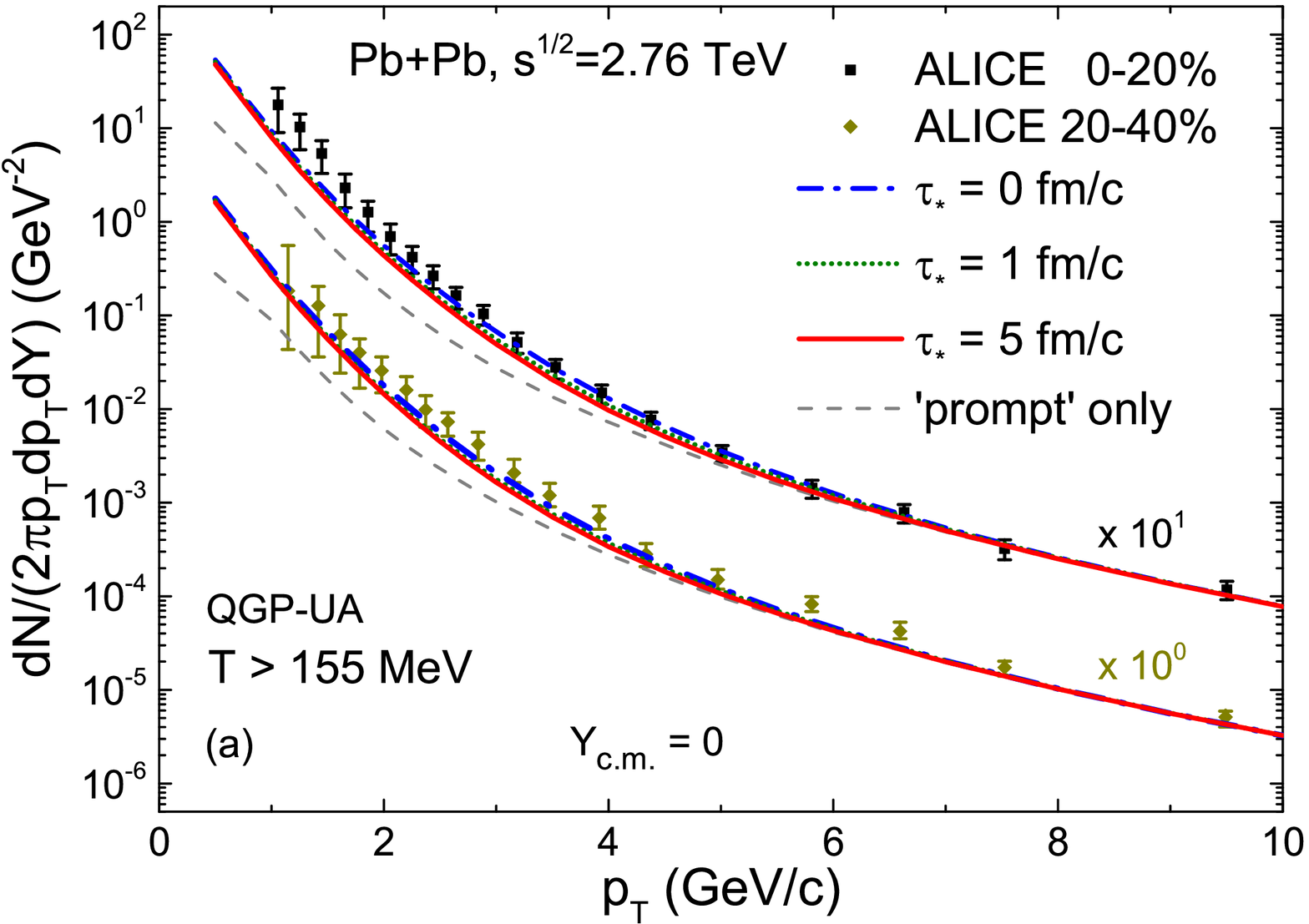}
\includegraphics[width=0.65\textwidth]{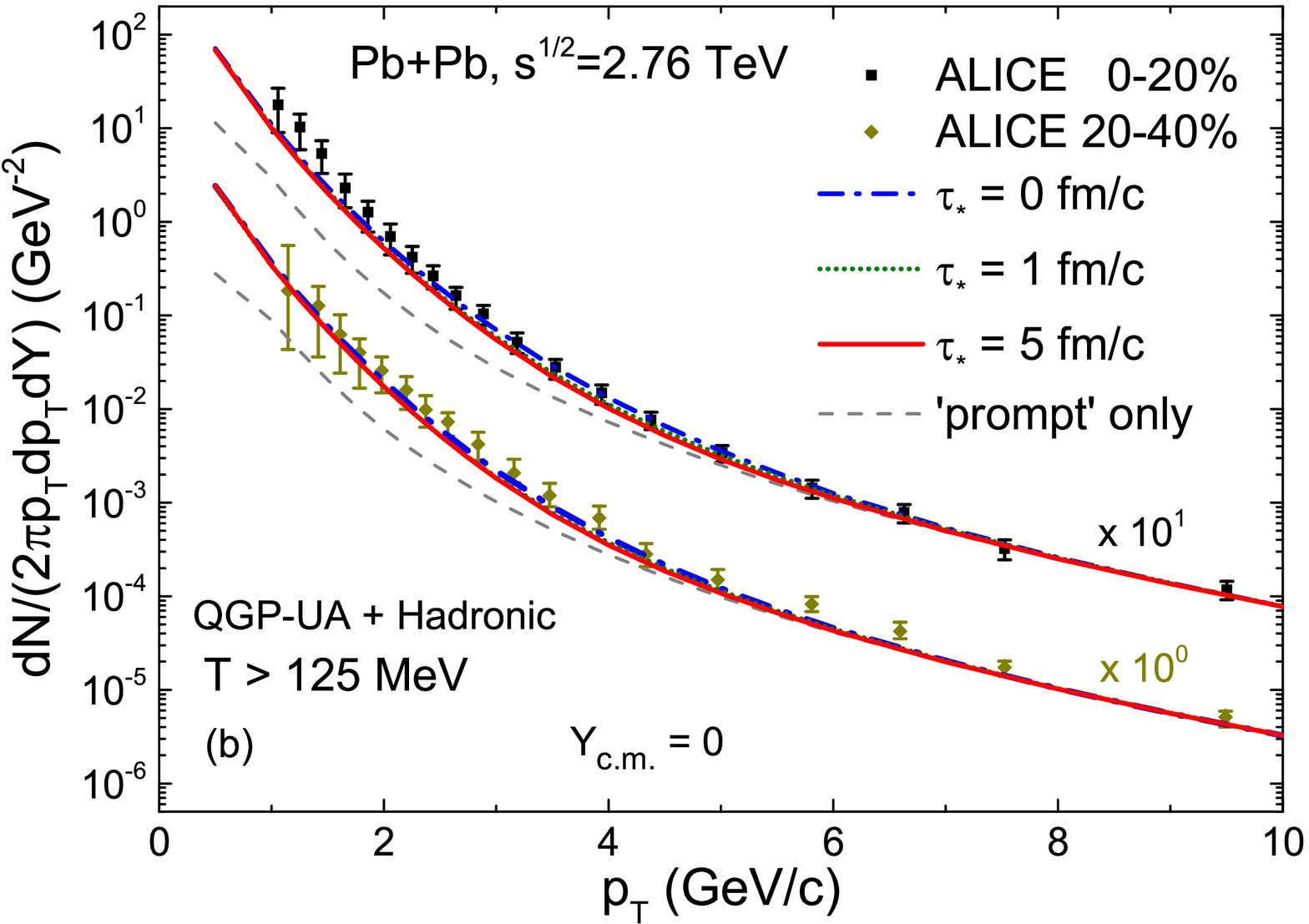}
\caption[]{(Color online)
Spectra of direct photons in {the 0--20\hspm\%}
central Pb+Pb collisions at $\sqrt{s_{NN}}=2.76~\textrm{TeV}$ calculated 
by using 
Eq.~(\ref{photsp})
with the cutoff temperatures \mbox{$T_f=155$}~(a) and~$125~\textrm{(b) MeV}$. 
At $T>155$~MeV the uQGP photon production rate given by Eq.~(\ref{uqgp2}) is used while for lower temperatures the hadronic PPR is employed.
The dash-dotted, dotted and solid curves
correspond to $\tau_*=0\hspm , 1$ and $5~\textrm{fm}/c$\,, respectively. Dots with error bars show the experimental
data~\cite{ALICEphot}.
}
\label{fig:phspec}
\end{figure*}
%--------------------------------------------------------------------

As mentioned above, spectra of direct photons include, in addition
to the thermal component, also the contribution of prompt photons
from initial collisions of nucleons in cold initial nuclei. This contribution is usually obtained
by using the perturbative QCD calculations of photon production
in a single $pp$\hsp -\hspm collision at the same
$\sqrt{s_{NN}}$. The obtained photon yield is scaled
by the average number of nucleon collisions for a given centrality class. Below we use the prompt photon spectra
in central Pb+Pb collisions at LHC reported in Refs.~\cite{Loh13,ALICEphot}. According to
our calculations, the contribution of prompt photons in such reactions becomes dominant
at high transverse momenta $p_{\hsp T}\gtrsim 5~\textrm{GeV}/c$\hsp . Unfortunately, this greatly reduces
the sensitivity of combined thermal and prompt photon \mbox{$p_{\hsp T}$--spectra} to the EoS and
to parameters of chemical nonequilibrium.

Figure~\ref{fig:phspec}\hsp{a} % and b 
shows our results for the direct photon spectrum in {the 0--20\hspm\%}
central Pb+Pb collisions at $\sqrt{s_{NN}}=2.76~\textrm{TeV}$,
calculated using the PPR from Eqs.~\eqref{uqgp1} and \eqref{uqgp2} 
with the
cut-off temperature of $T_f=155$~MeV.
We have checked that the
LA parametrization of thermal photon emission{, \re{uqgp1},} gives only several
percent lower yields as compared to the alternative UA choice. Therefore, we present here
only the results based on~\re{uqgp2}. 
To estimate contributions of thermal photon
emission from the late (hadronic) stages of the reaction, 
we additionally perform calculations for
the lower cut-off temperature of $125~\textrm{MeV}$, shown
in Fig.~\ref{fig:phspec}\hsp{b}.
In this case, for temperatures $T<155$~MeV
we use the parametrized PPR in the hadronic phase,
which includes contributions of the the in-medium $\rho$ mesons~\cite{Heffernan2015},
strange mesons~\cite{Turbide2004}, and the $\pi\pi$-bremsstrahlung~\cite{Heffernan2015}.
The consistency of our approach is provided by the fact that hadronic and QGP rates are
very similar in the vicinity of the crossover temperature,
as demonstrated in~\cite{Kap91,Paq15}.
In our calculations we use the chemically equilibrium rates
of photon emission from the confined phase.
In principle, the hadronic densities may be reduced due to the suppression
of constituent (anti)quarks with fugacities $\lambda<1$ (see the related discussion in Ref.~\cite{Vov16}). However, as one can see in Fig.~\ref{fig:T},
at these low temperatures the $\lambda$-values are already close to unity.
Having in mind that introducing the suppressed hadronic rates will require additional assumptions and that its effect is expected to be small, we do not implement the corresponding modification of hadronic PPR in the present paper.
Comparison of  Figs.~\ref{fig:phspec}\hsp{a and b}
shows that thermal photon emission from the low-temperature 
stage $T\lesssim 155~\textrm{MeV}$ gives only a slight
change of the yield at intermediate $p_{\hsp T}=2\div 6$~GeV/c.
On the other hand, such emission more noticeably increases the photon yield
at $p_{\hsp T}\lesssim 2~\textrm{GeV}/c$\hsp .
Including this additional hadronic source of thermal photons leads to a somewhat
better agreement with the observed data.
We note that PPR from hadronic phase are presently not constrained very well.
Thus, additional studies in that direction are needed.

The direct photon production in Pb+Pb collisions at LHC has been considered in various theoretical models which include relativistic ideal~\cite{Chatterjee12,Hees15} or viscous~\cite{Paq15} hydrodynamics, and the PHSD off-shell transport approach~\cite{Lin15}.
These studies describe experimental $p_T$-spectra of photons with a similar quality. Thus,
as noted in Ref.~\cite{ALICEphot}, the present uncertainties in the ALICE photon data do not allow to discriminate between various models and scenarios.

\begin{figure*}[hbt!]
\centering
\includegraphics[width=0.49\textwidth]{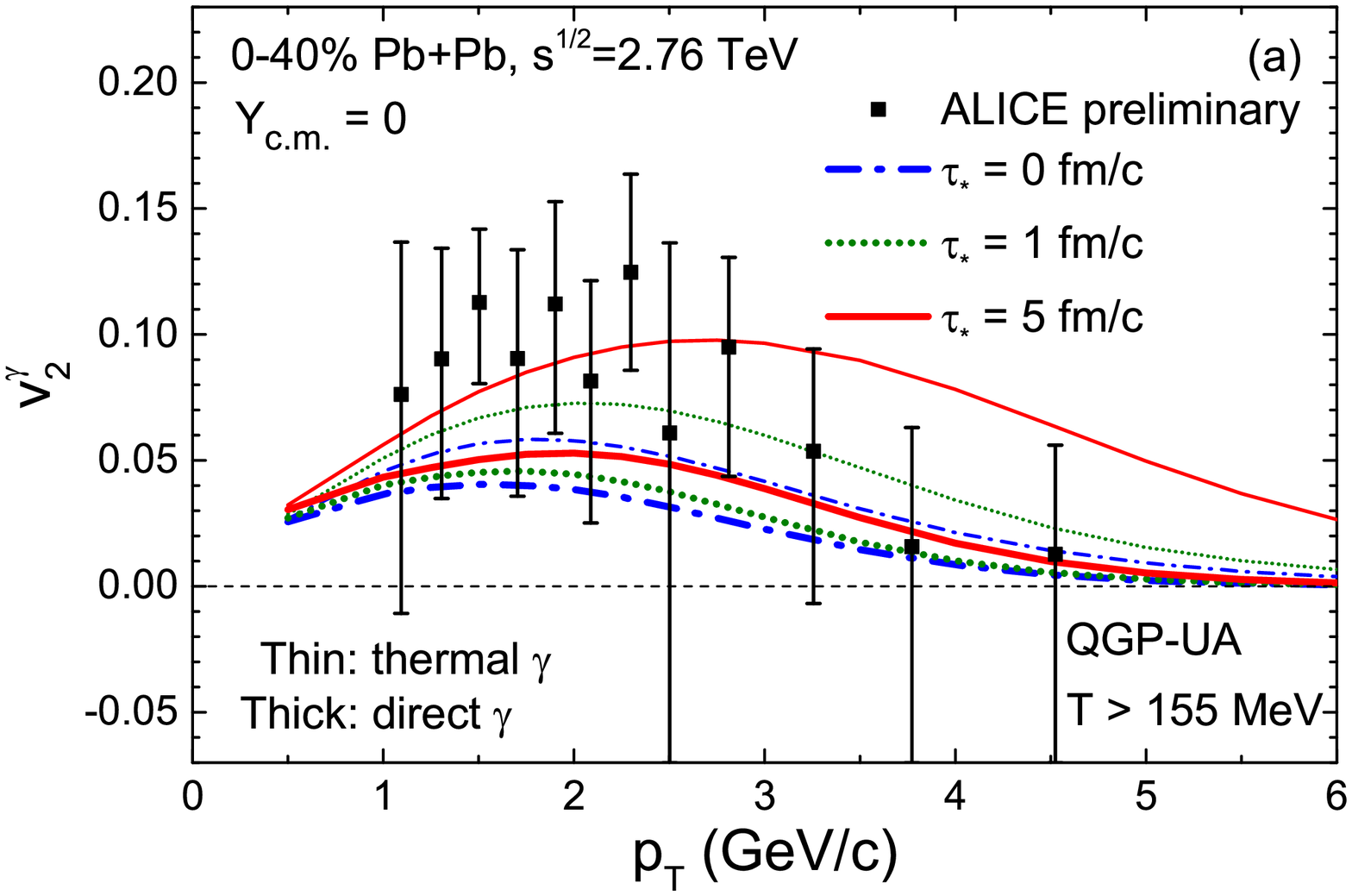}
\includegraphics[width=0.49\textwidth]{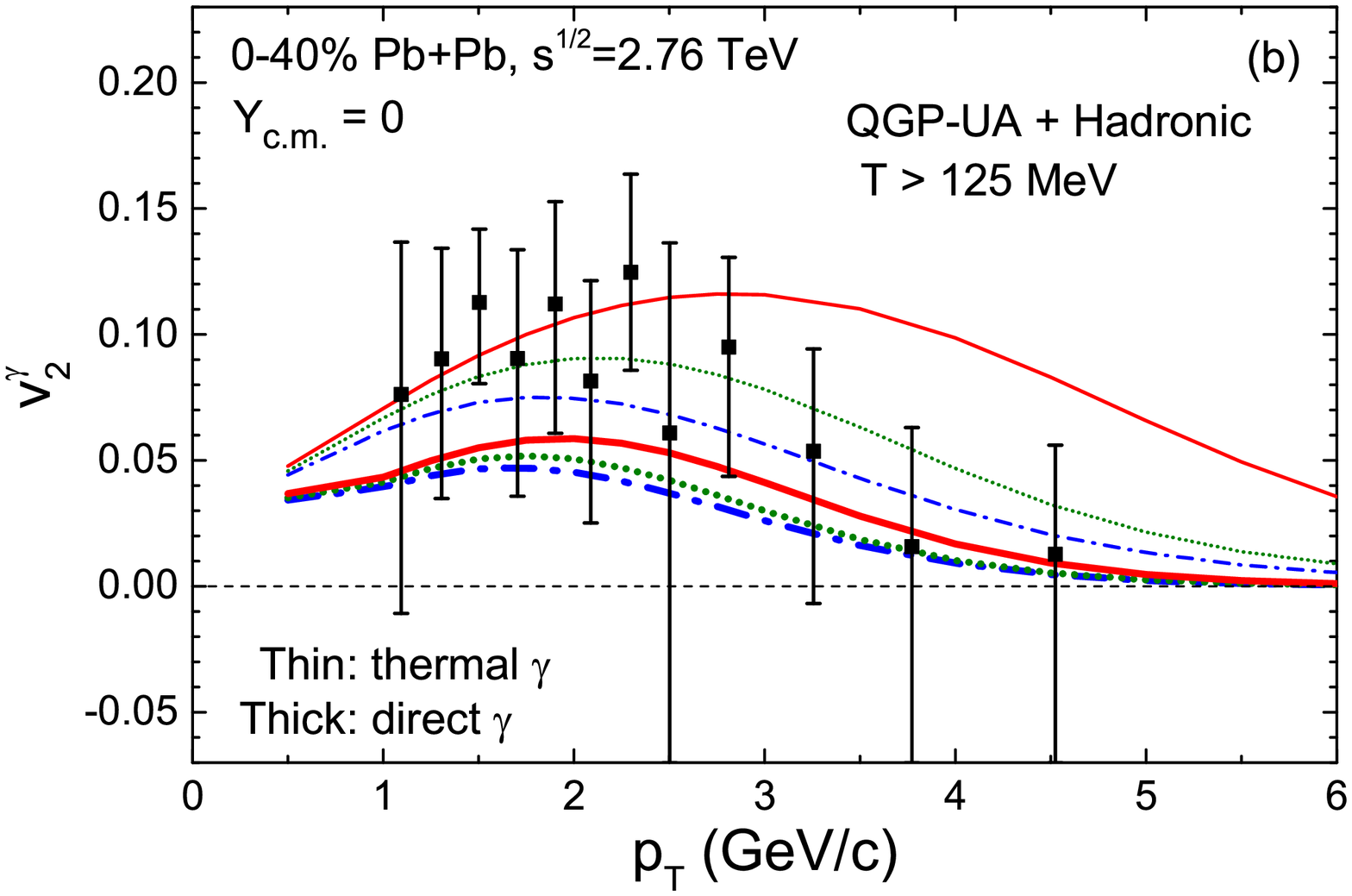}
\caption[]{(Color online)
Elliptic flow $v_{\hspm 2}^{\gamma}$ of direct photons as a function of transverse momentum $p_{\hsp T}$
in the~~$0-40$\hspm\%
central Pb+Pb collisions at $\sqrt{s_{NN}}=2.76~\textrm{TeV}$ calculated
with the cutoff temperatures \mbox{$T_f=155$}~(a) and~$125~\textrm{(b) MeV}$.
The dash-dotted, dotted and solid lines
correspond to $\tau_*=0\hsp , 1$ and $5~\textrm{fm}/c$\,, respectively\hspm .
Thick (thin) curves are calculated with (without) the contribution
of prompt photons in~\re{elfl}. Experimental data are taken from Ref.~\cite{Loh13}.
}
\label{fig:phv2}
\end{figure*}
%--------------------------------------------------------------------
The photon elliptic flow
%coefficient
$v_{\hspm 2}^{\gamma}(p_{\hsp T})$ is calculated by
\bel{elfl}
v_{\hspm 2}^{\gamma}(p_{\hsp T}) = \frac{\int_0^{2\hspm\pi} d\hsp\varphi \,
\frac{d N_{\gamma}}{d^{2} p_{\hsp T} dY} \, \cos(2\varphi)}{\int_0^{2 \pi} d\hsp\varphi\,\frac{d N_{\gamma}}{d^{2} p_{\hsp T} dY}}\,.
\ee
The photon spectrum, entering this equation includes both thermal and prompt components. We assume that prompt photons are azimuthally symmetric. Therefore, they contribute only to the denominator of~\re{elfl} reducing
$v_{\hspm 2}^{\gamma}$ at large $p_{\hsp T}$.

\vspace*{2mm}
\begin{figure}[htb!]
\centering
\includegraphics[width=0.99\textwidth]{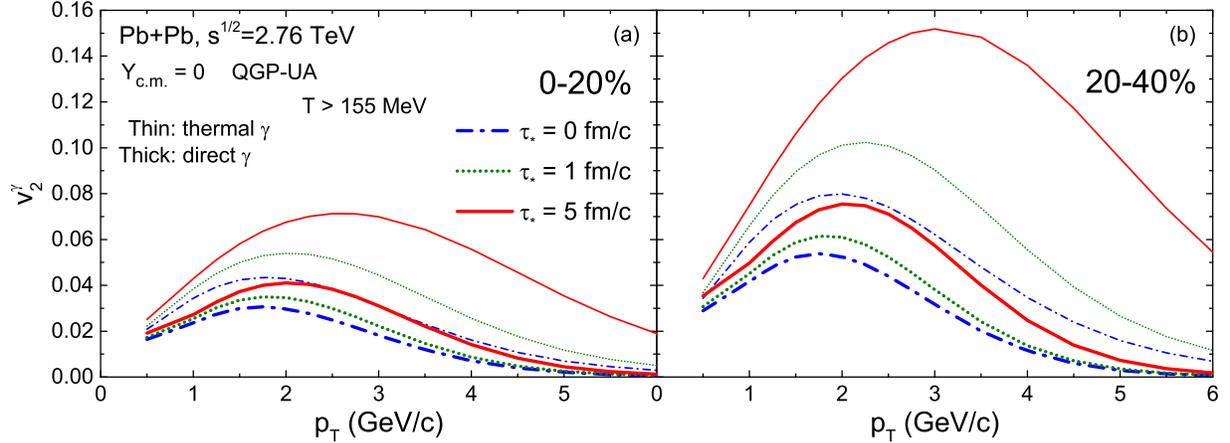}
\caption[]{(Color online)
Elliptic flow of the direct (thick lines) and thermal (thin lines) photons
for the 0--20\hspm\% (a) and 20--40\hspm\% (b) central Pb+Pb collisions at $\sqrt{s_{_{\rm NN}}} = 2.76$ TeV.
}\label{fig:ph-v2a}
\end{figure}
The results for direct photon elliptic flow in {the 0--40\hspm\%}
central Pb+Pb collisions at $\sqrt{s_{NN}}=2.76~\textrm{TeV}$ are shown in
Figs.~\ref{fig:phv2}\,{a} and b. One can see that the initial undersaturation of quarks
leads to a noticeable enhancement of $v_{\hspm 2}^{\gamma}$. The comparison of
thick and thin lines shows that this enhancement is significantly
reduced due to the presence of prompt photons\hspm\footnote
{
Similar conclusions have been made in Refs.~\cite{Liu14,Mon14}.
}.
From the analysis of different cutoff temperatures we conclude that the contribution of a colder `hadronic' stage increases the photon elliptic flow and leads to a somewhat better agreement with the ALICE data.
The latter ones are noticeably underestimated in the chemically equilibrium
scenario ($\tau_*=0$). The physical reason for the $v_2^{\gamma}$ increase for lower $T_f$ is rather clear.
It is explained by the increase of collective flow velocities at later times. Despite the fact
that fewer quarks are produced at late stages, their angular anisotropy will be stronger.

To study possible influence of the centrality choice, in Fig.~\ref{fig:ph-v2a}
we show the photon elliptic flows for the same reaction, but taking narrower centrality classes, 0--20\hsp\% and \mbox{20--40\hsp\%}.
One can see that the photon elliptic flow and its sensitivity to chemical nonequi\-librium effects becomes stronger for larger impact parameters. This behavior is explained by increased eccentricities of quark fireballs in less central events.

\section{Thermal dilepton emission}

The rate of thermal dilepton production from the lowest-order quark-antiquark
annihilation processes $q\hspm\ov{q}\to e^+e^-$ in the net baryon-free uQGP can be written as\hsp\footnote
{
An analogous expression in the limit of chemically equilibrated plasma~($\lambda=1$) has
been suggested in~\cite{Kaj86}. First calculations of the dilepton emission in uQGP have been presented in~Refs.~\cite{Str94,BK2}.
}:
\bel{drate}
\frac{dN}{d^{\hsp 4}x\hsp d^{\hsp 4}Q}=C_q\,\lambda^2\hsp\exp\left(-\,\frac{Q\hspm u}{T}\right),
\ee
where \mbox{$Q=p_+\hsp +\hsp p_-$} is the dilepton total four-momentum, and $T$ and $u$ are,
respectively, the local values of temperature and four-velocity of the medium. The coefficient in front of the rate is
$C_q=\frac{\ds\alpha^2}{\ds 4\pi^4}\hsp F_q$, where $\alpha$ and~$F_q$ are defined in \re{rat12}.
Note that~\re{drate} is obtained in the Boltzmann approximation for the (anti)quark
phase-space distributions and neglects the quark and lepton masses.
The $\lambda^2$ factor in~\re{drate} takes
into account the (anti)quark suppression in the chemically nonequilibrated QGP.

Introducing the dilepton invariant mass $M=\sqrt{Q^{\hsp 2}}$, one has
\bel{qcomp}
Q^{\,\mu}=(M_\perp\cosh{Y},\bm{Q}_\perp,M_\perp\sinh{Y})~,
\ee
where $M_\perp=\sqrt{M^2+\bm{Q}^2_\perp}$ stands for the transverse pair mass, and $Y=\tanh^{-1}(Q_z/Q_0)$ is the
longitudinal rapidity of the lepton pair.
Using {\re{boost}} for the four-velocity of the fluid in the (2+1)--dimensional hydrodynamics,
one gets the expression for the rest-frame dilepton's total energy
\bel{rfte}
(Q\hspm u)=\gamma_\perp\left[\hspm M_\perp\cosh\hsp (Y-\eta)-\bm{Q}_\perp \bm{v}_\perp\hsp\right].
\ee
Let us denote by~$\varphi$ and $\varphi_u$ the angles of $\bm{Q}_\perp$ and $\bm{v}_\perp$ with respect to
the $x$-axis, respectively. Then one can substitute
\mbox{$\bm{Q}_\perp \bm{v}_\perp=Q_\perp v_\perp\cos\hsp(\varphi-\varphi_u)$} in the right-hand side of~\re{rfte}.

From~\re{drate}, using the relations
\mbox{$d^{\,4}x=\tau d\tau\hsp d^{\,2}\hspace*{-1pt}x_\perp d\eta$} and
\mbox{$d^{\hsp 4}Q=M dM dY d^{\hsp 2}Q_\perp$}, after integrating
over the space--time rapidity $\eta$, we obtain
\bel{specd}
\frac{dN}{dM^{\hspm 2}dY d\hspm\varphi}=C_q\int\hspace*{-1pt}d^{\hsp 2}x_\perp\hspace*{-3pt}
\int\limits_{\tau_0}^{+\infty}\hspace*{-3pt}d\hspm\tau\hsp\tau\hsp
\lambda^2(\tau,\bm{x}_\perp)\,J(M,\tau,\bm{x}_\perp)\,\theta\hsp (T-T_f)\,,
\ee
{where}
\bel{specd1}
J(M,\tau,\bm{x}_\perp)=\int\limits_0^\infty\hspace*{-1pt}d\hspm Q_\perp Q_\perp
K_0\left(\frac{\gamma_\perp M_\perp}{T}\right)\exp{\left(\frac{\gamma_\perp \bm{Q}_\perp
\bm{v}_\perp}{T}\right)}.
\ee
Hereinafter we denote by $K_\nu(x)$ and $I_\nu(x)$ the modified Bessel functions of the order $\nu$.
Due to the assumed boost invariance, the dilepton spectrum (\ref{specd}) does not depend on $Y$ within the
(2+1)--dimensional hydrodynamics. Thus,
it should be applied essentially just in the central rapidity region.

Explicit relations for the invariant mass distribution and the elliptic flow
of thermal dileptons, obtained from~(\ref{specd}) and (\ref{specd1}), are
given in Eqs.~(\ref{spec2})--(\ref{v2d2}) of Appendix~B.
We would like to emphasize the well known fact that the dilepton mass spectrum does not depend explicitly
on the transverse collective velocity $\bm{v}_\perp$ {(see (\ref{spec2}) and (\ref{spec3}))}.
This is to be contrasted with the $p_{\hsp T}$-spectra
of thermal photons which are given by a superposition of exponents
$\exp{(-p_{\hsp T}/T_{\rm eff})}$, where~$T_{\rm eff}$ is the ''blue-shifted'' effective
tempera\-ture~$T_{\rm eff}=T\sqrt{(1+v_\perp)/(1-v_\perp)}$\,.

\vspace*{2mm}
\begin{figure}[htb!]
\centering
\includegraphics[width=0.99\textwidth]{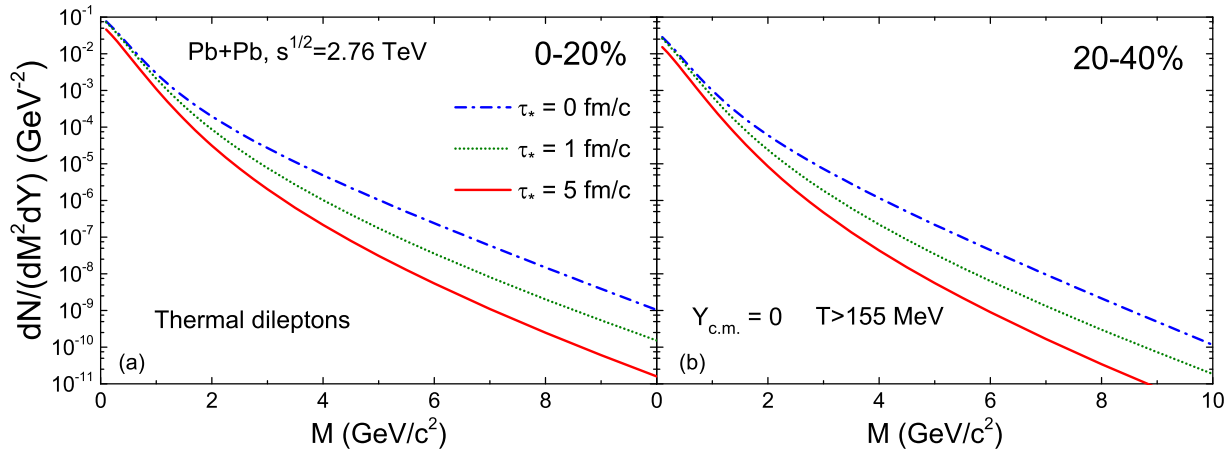}
\caption[]{(Color online)
Invariant mass distribution of thermal dileptons in {the 0--20\hspm\%~(a)
and 20--40\hspm\%~(b)} central Pb+Pb collisions at $\sqrt{s_{NN}}=2.76~\textrm{TeV}$ calculated
for $\tau_* = 0, 1~\textrm{and}~5~\textrm{fm}/c$\,. {All results correspond to}
the cut-off temperature $T_f=155~\textrm{MeV}$.
}
\label{fig:dilept-dNdM-155}
\end{figure}
In the limiting case of the one--dimensional Bjorken-like hydrodynamics one gets for purely central collisions~\cite{Sto15}
\bel{dspeb}
\frac{\ds dN}{\ds dM^{\hsp 2}dY}\simeq 2\hspm\pi^2 R^{\hsp 2} C_q M
\hspace*{-2pt}\int\limits_{\tau_0}^{\tau_f}\hspace*{-3pt}
d\hspm\tau\hsp\tau\,T(\tau)\hsp K_1\hspace*{-2pt}\left[\frac{M}{T(\tau)}\right]\hspace*{-2pt}\lambda^2(\tau)\,
~~~~~(v_\perp=0)\,,
\ee
where $R$ is the geometrical radius of colliding nuclei and $\tau_f$ is determined from $T(\tau_f)=T_f$.

The results of calculating the dilepton mass spectrum in
central Pb+Pb collisions at \mbox{$\sqrt{s_{NN}} = 2.76$~TeV} are shown
in~Fig.~\ref{fig:dilept-dNdM-155}
for the cut-off temperature $T_f=155~\textrm{MeV}$.  One can see that the initial
quark suppression leads to a strong reduction of the dilepton yield at $M\gtrsim 2~\textrm{GeV}$.
Note that we do not include contributions of hard (Drell-Yan) dileptons~\cite{Lin16}  produced in binary collisions of initial nucleons.

\begin{figure}[htb!]
\centering
\includegraphics[width=0.99\textwidth]{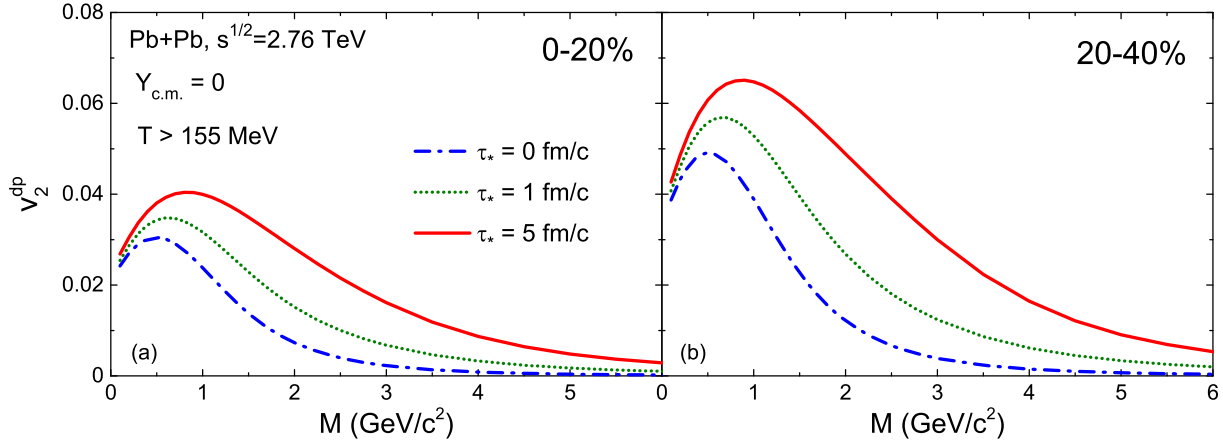}
\caption[]{(Color online)
Same as Fig.~\ref{fig:dilept-dNdM-155} but for elliptic flow
of thermal dileptons $v_2^{\rm dp}$.
}\label{fig:dilept-v2}
\end{figure}
{As shown in Appendix B}, the elliptic flow of {thermal} dileptons strongly
depends both on the direction and magnitude of the transverse {collective} velocity.
Note that the previous analysis of dilepton elliptic flow of Refs.~\cite{Cha07,Vuj16}
corresponds to the limit of complete chemical equilibrium ($\lambda=1$).

The elliptic flows of thermal dileptons in the same reaction, calculated in accordance with \re{v2d1},
are shown in Fig.~\ref{fig:dilept-v2} for several values of $\tau_*$. Similar to direct photons we predict
a strong enhancement of the dilepton elliptic flow
as compared to the equilibrium sce\-nario~($\tau_*=0$). Note that $v^{\rm dp}_2$ values are larger
for more peripheral events.

\section{Conclusions}

In this paper we presented calculations of the electromagnetic observables in Pb+Pb collisions at LHC energies
for different assumptions about the initial state of
produced partonic matter. In our calculations we have used a rather
advanced hydrodynamic model which was previously used to describe 
%successfully 
the hadron observables.

In the non-equilibrium scenario, the thermal production of high-$p_{\hsp T}$ photons is significantly suppressed compared to the equilibrium case.
However, since the high-$p_{\hsp T}$ %part of the direct photons spectrum
photon production
is dominated by the %pQCD processes in the initial stage of the collision
prompt photons from initial parton scatterings,
we
do not find a strong suppression
of the total direct photon spectra.
Our analysis shows that the
$p_{\hsp T}$-spectra of such photons calculated for equilibrium and nonequilibrium
scenarios differ at most by a factor of 2, and these differences are within the error
bars of experimental data.
Much stronger effects are found for the thermal dilepton spectra, especially
at large invariant masses $M\gtrsim 2~\textrm{GeV}$, where the deviations between two
scenarios can reach one to two orders of magnitude. Unfortunately, the corresponding experimental
data are not available yet.

Our hydrodynamic approach also allows us to calculate the elliptic flow
parameters~$v_2^{\gamma, dp}$, which characterize the azimuthal anisotropy of the direct photon and dilepton
emission. We find a rather significant enhancement of the elliptic flow for the pure glue initial
state for both photons and dileptons.
However, the available experimental data
for photons are not yet accurate enough to discriminate between the considered scenarios
for the initial stage.
We are looking forward 
for the more precise experimental
data enabling more definite statements on the
evolution of primordial matter in ultrarelativistic heavy-ion collisions.

\begin{acknowledgments}
The authors thank E. L. Bratkovskaya, C. Gale, H. Niemi, L. Pang, and M. Strickland for useful discussions and fruitful comments.
This work was partially supported by the Helmholtz International Center for
FAIR, Germany, and by the Goal-Oriented Program of the National Academy of Sciences of Ukraine and the European Organization for Nuclear Research (CERN), Grant CO-1-3-2016. L.M.S. and I.N.M. acknowledge a partial support from
the grant NSH-932.2014.2 of the Russian Ministry of Education and Science.
\end{acknowledgments}

\appendix
\section{Photon Emission}
\label{app-A}
{In this section we consider the PPR in the} chemically equilibrated QGP.
Processes 1 and 2 {(see Sec.~\ref{dpes})} have been analyzed in Ref.~\cite{Kap91}. The infrared divergencies of photon
production cross sections were regularized by using the hard thermal loop resummation procedure~\cite{Pis88}.
The following expressions for invariant rates of photon {production} have been obtained
in the lowest order approximation {in the strong coupling constant $\alpha_{\ds s}$}:
\bel{rat12}
\Gamma_i(\widetilde{E},T)\equiv E\frac{\ds dN_{\hspm i}}
{\ds d^{\hsp 3}\hspace*{-1pt}p\hsp d^{\hsp 4}\hspace*{-1pt}x}=
A_{\hspm i}\hspm F_q\hsp\alpha\hspm\alpha_{\ds s} T^2e^{-\ds x}\ln\frac{B_{\hsp i}\hsp x}{\alpha_s}~~~(i=1,2)\hsp.
\ee
Here $p^{\,\mu}=(E,\bm{p})^{\mu}$ is the photon four-momentum, $\widetilde{E}=p_{\mu}\hsp u^{\mu}$ is the
{rest-frame} photon energy, \mbox{$\alpha=e^{\hsp 2}\simeq 1/137$} is the electromagnetic coupling constant,
\mbox{$F_q=\sum_f\left(\frac{\ds e_f}{\ds e}\right)^2$} ($e_f$ is the charge of quarks with
flavor~$f$)\,, and $x=\widetilde{E}/T$.
Numerical values of constants~$A_{\hspm i}, B_{\hsp i}$ are given by the relations
\bel{cexp}
A_2=2\hspm A_1=\frac{\ds 1}{\ds 3\hspm\pi^2},~~B_1\simeq 1.00,~~B_2\simeq 0.112\hspm .
\ee
 In the following we assume the number of quark flavours $N_f=3$ and take into account
 the temperature dependence of $\alpha_{\ds s}$ by using the parametrization~\cite{Ste01}
 \bel{scc}
 \alpha_{\ds s}=\frac{\ds 6\pi}{\ds (33-2N_f)\ln{(8\hsp T/T_*)}}\,,
 \ee
 where $T_*=170~\textrm{MeV}$.

%-------------------------------------------
     \begin{figure*}[hbt!]
%\centerline{\includegraphics[trim=0 7.5cm 0 8.5cm, clip, width=0.8\textwidth]{e-gam-018}}
\centerline{\includegraphics[width=0.7\textwidth]{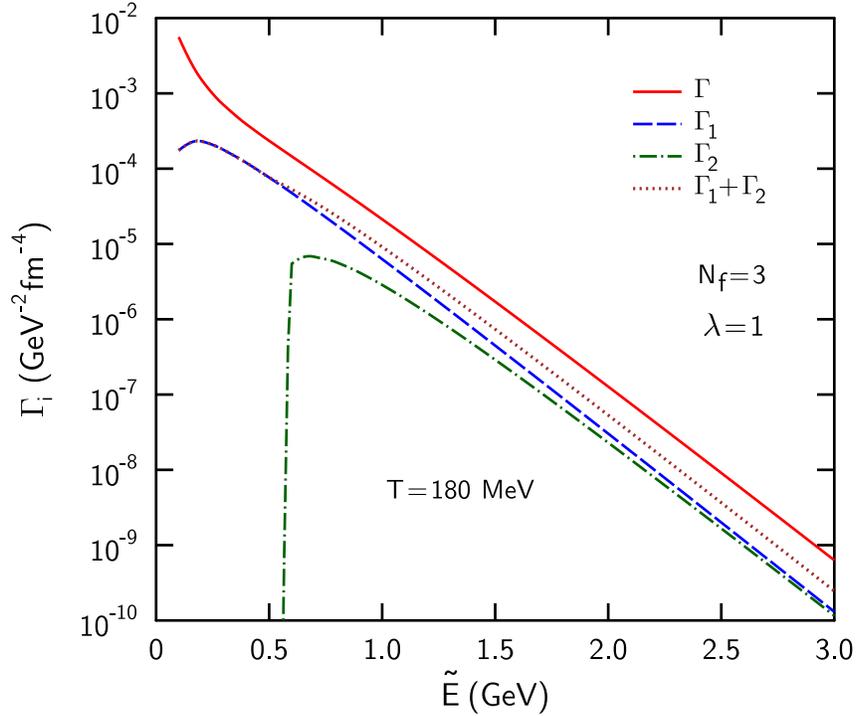}}
        \caption{(Color online)
         {Thermal} photon production rates in equilibrium QGP as functions of the rest-frame photon energy at temperature $T=180$ MeV.}
        \label{fig:e-ppr}
        \end{figure*}
%---------------------------------------------
Processes 3 and 4 correspond to higher orders in $\alpha_{\ds s}$. The detailed calculations in~\cite{Arn01}
give the following result for the total PPR:
\begin{eqnarray}
&&\Gamma(\widetilde{E},T)=\sum_{i=1}^{4}\Gamma_i(\widetilde{E},T)=\frac{1}{\pi^2}\hsp F_q\hsp\alpha\hspm\alpha_{\ds s} T^2\Phi (x)\,,\label{ramy1}\\
&&\Phi(x)=(e^x+1)^{-1}\left[\frac{1}{2}\hsp\ln{\frac{\ds 3\hsp x}{\ds 2\hspm\pi\alpha_{\ds s}}}
+C_{12}(x)+C_{34}(x)\right],\label{ramy2}
\end{eqnarray}
where
\bel{ramy3}
C_{12}(x)=0.041\hsp x^{-1}-0.3615+1.01\hsp e^{-1.35\hsp x},
\ee
and
\bel{ramy4}
C_{34}(x)=\sqrt{1+\frac{N_f}{6}}\left[\frac{0.548}{x^{3/2}}\ln{(12.28+x^{-1})}+\frac{0.133\hsp x}{\sqrt{1+x/16.27}}\right].
\ee
These formulas become not accurate outside the domain $0.2\lesssim x\lesssim 50$\,.
Figure~\ref{fig:e-ppr} shows numerical values of  $\Gamma_1, \Gamma_2$, and $\Gamma$ for $T=180~\textrm{MeV}$.
One can see that contributions of processes 3 and 4 are rather significant at all considered
photon energies.

\section{Dilepton Emission}
\label{app-B}

The invariant mass spectrum $dN/dM^{\hspm 2}dY$ and the elliptic flow $v_2=v_2(M)$ of thermal dileptons are determined
by integrating \re{specd} over $\varphi$ with the weights $1$ and $\cos{(2\hspm\varphi)}$\hsp, respectively.
We~get the relations
\begin{eqnarray}
\frac{\ds dN}{\ds dM^{\hspm 2}dY}&=&C_q\int\hspace*{-1pt}d^{\hsp 2}x_\perp\hspace*{-3pt}
\int\limits_{\tau_0}^{+\infty}\hspace*{-3pt}d\hspm\tau\hsp\tau \lambda^2(\tau,\bm{x}_\perp)\,
J_1(M,\tau,\bm{x}_\perp)\,\theta\hsp (T-T_f),\label{spec2}\\
v_2\frac{\ds dN}{\ds dM^{\hspm 2}dY}&=&C_q\int\hspace*{-1pt}d^{\hsp 2}x_\perp\hspace*{-3pt}
\int\limits_{\tau_0}^{+\infty}\hspace*{-3pt}d\hspm\tau\hsp\tau  \lambda^2(\tau,\bm{x}_\perp)\,
J_2(M,\tau,\bm{x}_\perp)\,\theta\hsp (T-T_f),\label{v2d1}
\end{eqnarray}
where
\bel{spec3}
J_1=\int\limits_0^{2\pi} d\hspm\varphi\,J=2\pi\int\limits_0^\infty\hspace*{-1pt}d\hspm
Q_\perp Q_\perp K_0\left(\frac{\gamma_\perp M_\perp}{T}\right)\,I_0\left(\frac{\gamma_\perp v_\perp Q_\perp}{T}\right)=2\pi M\hsp T\hsp K_1\left(\frac{M}{T}\right)
\ee
and
\begin{eqnarray}
\nonumber
J_2=\int\limits_0^{2\pi} d\hspm\varphi\,J\hsp \cos{(2\hspm\varphi)}&=&2\pi\cos{(2\hspm\varphi_u)}
\int\limits_0^\infty d\hspm Q_\perp Q_\perp K_0\left(\frac{\gamma_\perp M_\perp}{T}\right)\,I_2\left(\frac{\gamma_\perp v_\perp Q_\perp}{T}\right)\\
&=&\cos{(2\hspm\varphi_u)}\left\{J_1-\frac{4\pi T^{\hsp 2}}{\gamma_\perp^2-1}
\left[K_0\left(\frac{M}{T}\right)-K_0\left(\frac{\gamma_\perp  M}{T}\right)\right]\right\}.
\label{v2d2}
\end{eqnarray}

To calculate integrals over $Q_\perp$ in Eqs.~(\ref{spec3}) and (\ref{v2d2}),
we have applied a procedure suggested in~\cite{Wat95}. We start from the integral representation
\bel{beir}
K_\nu(x)=x^\nu\int\limits_0^\infty\frac{dt}{t^{\nu+1}}
\exp{\left[-\frac{1}{2}\left(t+\frac{x^2}{t}\right)\right]},
\ee
and then use the formulas
\begin{eqnarray}
\int\limits_0^\infty\hspace*{-1pt}d\hspm Q_\perp Q_\perp e^{-AQ_\perp^{\hspm 2}}
I_0(B\hspm Q_\perp)&=&\frac{1}{2 A}\exp{\left(\frac{~B^{\hspace*{.2pt}2}}
{4 A}\right)}\,,\label{inti0}\\
\int\limits_0^\infty\hspace*{-1pt}d\hspm Q_\perp Q_\perp e^{-AQ_\perp^{\hspm 2}} I_2(B\hspm Q_\perp)&=&\left(\frac{1}{2A}-\frac{2}{~B^{\hspace*{.2pt}2}}\right)\hsp
\exp{\left(\frac{~B^{\hspace*{.2pt}2}}{4 A}\right)}+\frac{2}{B^{\hspace*{.3pt}2}}\,.\label{inti2}
\end{eqnarray}
The second equation is obtained by using the relation \mbox{$I_2(x)=I_0(x)-2\hspm I_0^{\prime}(x)/x$}.

Note that \mbox{$\cos{(2\varphi_u)}=(v_x^2-v_y^2)/v_\perp^2$ in~\re{v2d2}}. It is easy to show
that at small $v_\perp$ one gets the approximate relation
\bel{j2ap}
J_2\simeq \frac{\pi}{2}\,(v_x^2-v_y^2\hsp)\,M^{\hsp 2}K_2\left(\frac{M}{T}\right)~~~~(v_\perp\ll 1)\,.
\ee
Presumably, this limiting case corresponds to early stages of the QGP evolution.

\end{document}